\documentclass[bibyear]{aa}
\usepackage{amsmath,amstext}
\usepackage{natbib,twoopt}
\usepackage[breaklinks=true,hidelinks]{hyperref}
\usepackage{amssymb}
\usepackage{wasysym}

\def\spose#1{\hbox to 0pt{#1\hss}}
\def\lta{\mathrel{\spose{\lower 3pt\hbox{$\mathchar"218$}}
     \raise 2.0pt\hbox{$\mathchar"13C$}}}
\def\gta{\mathrel{\spose{\lower 3pt\hbox{$\mathchar"218$}}
     \raise 2.0pt\hbox{$\mathchar"13E$}}}

\begin{document}

\titlerunning{LRDs as obscured LBDs}

\title{Little red dots as obscured little blue dots: a super-Eddington unification model}

\author{Piero Madau\inst{1,2}
\and
Roberto Maiolino\inst{3,4,5}
}
\institute{Dipartimento di Fisica ``G. Occhialini,'' Università degli Studi di Milano-Bicocca, Piazza della Scienza 3, I-20126 Milano, Italy \and Department of Astronomy \& Astrophysics, University of California, 1156 High Street, Santa Cruz, CA 95064, USA \and
Kavli Institute for Cosmology, University of Cambridge, Madingley Road, Cambridge CB3 0HA, UK 
\and Cavendish Laboratory, University of Cambridge, 19 JJ Thomson Avenue, Cambridge CB3 0HE, UK \and
Department of Physics and Astronomy, University College London, Gower Street, London WC1E 6BT, UK}

\abstract{
We investigate whether Little Red Dots (LRDs) are the dust-reddened, high-inclination counterparts of compact, blue broad-line AGNs (``Little Blue Dots'', LBDs) powered by super-Eddington accretion. We model the central engine as a geometrically thick, radiation--pressure supported accretion flow whose funnel produces strongly anisotropic, intrinsically blue ionizing continua, coupled to an equatorially concentrated broad-line region (BLR) and dusty reprocessing clouds with modest covering factor.
Using inclination-dependent spectral energy distributions (SEDs) as input to \textsc{Cloudy}, we show that the extreme broad H$\alpha$ equivalent widths (EWs) of JWST LRDs can be reproduced with global BLR covering factors of only $C_{\rm BLR}\simeq 0.1$, fully consistent with standard Type~1 AGNs and far below unity. Large Balmer EWs arise because self-shadowing suppresses the high-inclination optical continuum while the BLR is illuminated by an ionizing-rich EUV SED. Weak high-ionization lines (e.g.\ He\textsc{ii}\,$\lambda4686$) follow from the orientation-dependent suppression of the XUV/soft X-ray continuum toward equatorial directions, without requiring a fully enclosing   gaseous ``cocoon''.
Applying a gray dust attenuation law with $A_V\simeq 3$ along high-inclination (LRD-selected) sightlines, our fiducial model reproduces the V-shaped UV-optical continua of LRDs and the large Balmer decrements; strong Balmer breaks arise only along the most obscured sightlines. A compact equatorial dust structure with modest global covering factor intercepts and reradiates only a small fraction of the bolometric luminosity, yielding a modest hot-dust bump and far-IR/sub-mm output consistent with current measurements and limits and implying small dust masses. This single-framework model links LRD and LBD observables through orientation, predicting correlated trends in H$\alpha$ EW, Balmer decrement, Balmer break, high-ionization line strengths, and IR emission.
}


\maketitle
\section{Introduction}
\label{sec:intro}

The James Webb Space Telescope (JWST) has uncovered a rich population of moderate-luminosity broad-line active galactic nuclei (BLAGNs) at $z \gtrsim 4$, powered by accretion onto early massive black holes (MBHs) with inferred masses $\sim 10^{6}$--$10^{8}\,M_\odot$ \citep[e.g.,][]{Harikane2023AGN,MaiolinoAGN,Taylor2025_BHMF,Juod2026}. 
These sources appear to outnumber the extrapolation of the quasar population at low luminosities by orders of magnitude, and constraints from existing multiwavelength data -- in particular deep X-ray imaging and stacking analyses --  indicate that many are unusually weak in X-rays compared to low-redshift Type~1 AGNs \citep[e.g.,][]{Ananna2024,Yue2024,Maiolino2025}.
Their space density, spectral shapes, and weak high-energy emission challenge standard inferences about MBH growth, accretion physics, and AGN selection in the first few Gyrs after the Big Bang.

A striking empirical aspect of this JWST BLAGN census is their compactness: many appear as unresolved or barely resolved ``little dots'' in rest-frame optical imaging. Recent work has proposed a useful shorthand that classifies these compact JWST BLAGNs according to their UV-optical continuum slopes into ``Little Red Dots'' (LRDs) and ``Little Blue Dots'' (LBDs), with LRDs denoting the red rest-optical, V-shaped minority subset and LBDs the comparably compact, bluer-continuum majority \citep{Brazzini2026,Geris2026}. {In JADES, LRD-like objects comprise approximately 20\% of the BLAGN sample, with their incidence showing a strong luminosity dependence \citep{Hainline2025, MadauMaiolinoLF}.}

LRDs are commonly identified as sources with a V-shaped UV-optical SED in which the continuum turns sharply at $\lambda_{\rm rest}\sim 0.35$--$0.40\,\mu$m, near the Balmer limit \citep[e.g.,][]{Matthee2024,Greene2024,Kocevski2025,Wang2025,Hainline2025,Akins2025a,Delvecchio2025,Barro2025}. Homogeneous spectroscopy has clarified that the LRD phenomenology is not defined by color alone: objects exhibiting a V-shaped continuum are strongly associated with broad Balmer emission and with a dominant unresolved component in the rest-frame optical \citep{Hviding2025}. This empirical triad -- V-shaped UV-optical continuum, broad Balmer lines, and a rest-optical point-source component -- suggests that the central engine contributes substantially to the observed light, even when broadband colors might otherwise encourage a stellar interpretation. Several of these characteristics -- enhanced Balmer emission, weak high-ionization features, and weak or absent X-rays -- extend into the broader BLAGN population, with LRDs occupying the most extreme tail \citep{Hviding2025}.

A key clue is that these JWST BLAGNs are extreme Balmer emitters while simultaneously showing weak high-ionization rest-UV features. In the RUBIES census, LRDs occupy a particularly extreme locus in the $L_{\mathrm{H}\alpha}$--$M_{\rm UV}$ plane: while LRDs are faint in the rest-UV at fixed $L_{\mathrm{H}\alpha}$, the most extreme H$\alpha$ emitters are dominated by LRDs, i.e. LRDs constitute the most luminous H$\alpha$ emitters at fixed UV luminosity \citep{Hviding2025}. Consistent with this, the median rest-frame EW reported for the broad H$\alpha$ line in JWST-identified $z>4$ BLAGNs is $\simeq 570$~\AA, roughly a factor of $\sim 3$ higher than in typical low-$z$ AGNs \citep{Maiolino2025}, while in the LRDs subpopulation the H$\alpha$ EW is even higher \citep{degraaf2025_multipleBHstars}. Yet high-ionization rest-UV lines such as C{\sc iv}, He{\sc ii}, and N{\sc v} are frequently weak or undetected in BLAGNs with rest-UV coverage, including the LRD subset \citep[e.g.,][]{Tang2025,Juod2026,Zucchi2026,Lambrides2026}. The joint occurrence of extreme Balmer emission and suppressed high-ionization lines places strong constraints on the ionizing SED incident on the BLR and on the geometry and obscuration of the line-emitting region. In anisotropic-emission geometries where the observer receives a continuum boosted relative to the radiation intercepted by the BLR, recombination-line EWs are expected to be reduced, making H$\alpha$ an orientation-sensitive diagnostic \citep[e.g.,][]{Madau2026}.

Although the origin of the LRD continuum remains debated, the spectroscopic and morphological evidence motivates an AGN-dominated interpretation for at least a substantial fraction of the population. Motivated by the emerging LBD/LRD taxonomy, we consider an orientation-obscuration continuity picture in which LRDs are not a separate class of engines, but rather represent LBD-like BLAGNs viewed through substantial circumnuclear dust and gas. In this view, LRDs and LBDs share a common engine, with both classes potentially subject to some host-galaxy reddening, while the V-shaped SED and red rest-optical slope of LRDs arise from additional circumnuclear attenuation and reprocessing along higher-inclination sightlines. Such a geometry naturally connects compactness, optical redness, weak variability, and suppressed X-rays to line-of-sight processing, while allowing broad lines to remain visible. {Recent analyses likewise suggest that LBDs and LRDs do not constitute two sharply distinct populations, but instead occupy opposite ends of a continuous sequence in compactness, continuum shape, and broad-line prominence \citep{Billand2026}.}

A natural physical framework that can supply both strong anisotropy and self-occultation is rapid accretion in the super-Eddington regime. Geometrically thick flows with funnel-like structures produce strong inclination-dependent SEDs and self-shadowing, which can reshape the UV-optical continuum and modulate the illumination of the BLR \citep{Wang2014,Lupi2024b,Madau2026}. In addition, X-ray emission can be intrinsically weak if a hot corona is efficiently Compton-cooled by the intense soft-photon field in the inner flow, suppressing hard X-rays even before line-of-sight obscuration is considered \citep{Madau2024,Trinca2026}. Here, we investigate whether the combination of dense BLR gas, anisotropic emission from super-Eddington accretion, and dust reddening can provide a coherent framework for the defining features of LRDs -- their V-shaped UV-optical continua, extreme Balmer emission (including large H$\alpha$ equivalent widths), weak high-ionization lines, and suppressed variability -- and we assess how LRDs relate to the broader population of JWST-discovered high-redshift BLAGNs, including the putative LBD majority.

\section{Super-Eddington accretion}

We model the engines of JWST BLAGNs at $z>4$ as geometrically thick, moderately super-Eddington accretion flows with rates $\dot m
\simeq 10$--$30$, where $\dot m\equiv \dot M/\dot M_{\rm Edd}$ and $\dot M_{\rm Edd}\equiv 10\,L_{\rm Edd}/c^2$.  
For these accretion rates, the corresponding Eddington ratio is $\lambda_{\rm Edd}\equiv L_{\rm bol}/L_{\rm Edd}\simeq 4$--$7$, reflecting the decline of the effective radiative efficiency at increasing $\dot m$ in thick-disk solutions, where the inner edge (cusp) shifts inward toward the marginally bound orbit, reducing the binding energy released as radiation \citep{Paczynsky1980}.
We adopt the non-advective, radiation-pressure supported torus formalism in which the disk surface follows equipotential contours set by an assumed specific-angular-momentum distribution \citep{Paczynsky1980}. On the photosphere, hydrostatic balance requires that the net outward radiative flux, $F_{\rm net}=F_{\rm out}-F_{\rm in}$, balances the local effective gravity \citep[e.g.,][]{Sikora1981,Madau1988}. Because the funnel walls are mutually visible, nonlocal self-irradiation implies $F_{\rm in}\neq 0$. In the idealized perfect-reflection limit used here, incident radiation is re-emitted without thermalization, so the local temperature is set by the locally generated flux while the outward flux $F_{\rm out}$ can be enhanced by reflected/scattered contributions. The resulting ``mirror'' funnel naturally yields self-consistent super-Eddington thick-disk geometries with strong collimation toward the rotation axis and self-shadowing at high inclinations
\citep{Sikora_W1981}.

Global MHD and radiation-GRMHD simulations of mildly supercritical accretion broadly support this qualitative picture, producing radiation-pressure supported inner tori, equatorial outflows, and a narrow funnel-shaped photosphere that channels escaping radiation toward the symmetry axis \citep[e.g.,][]{Jiang2014,Pacucci2024,ZhangGRMHD2025}.

\noindent
\begin{figure}[!htb]
\centering
\includegraphics[width=\hsize, trim=0 0 0 0,clip]{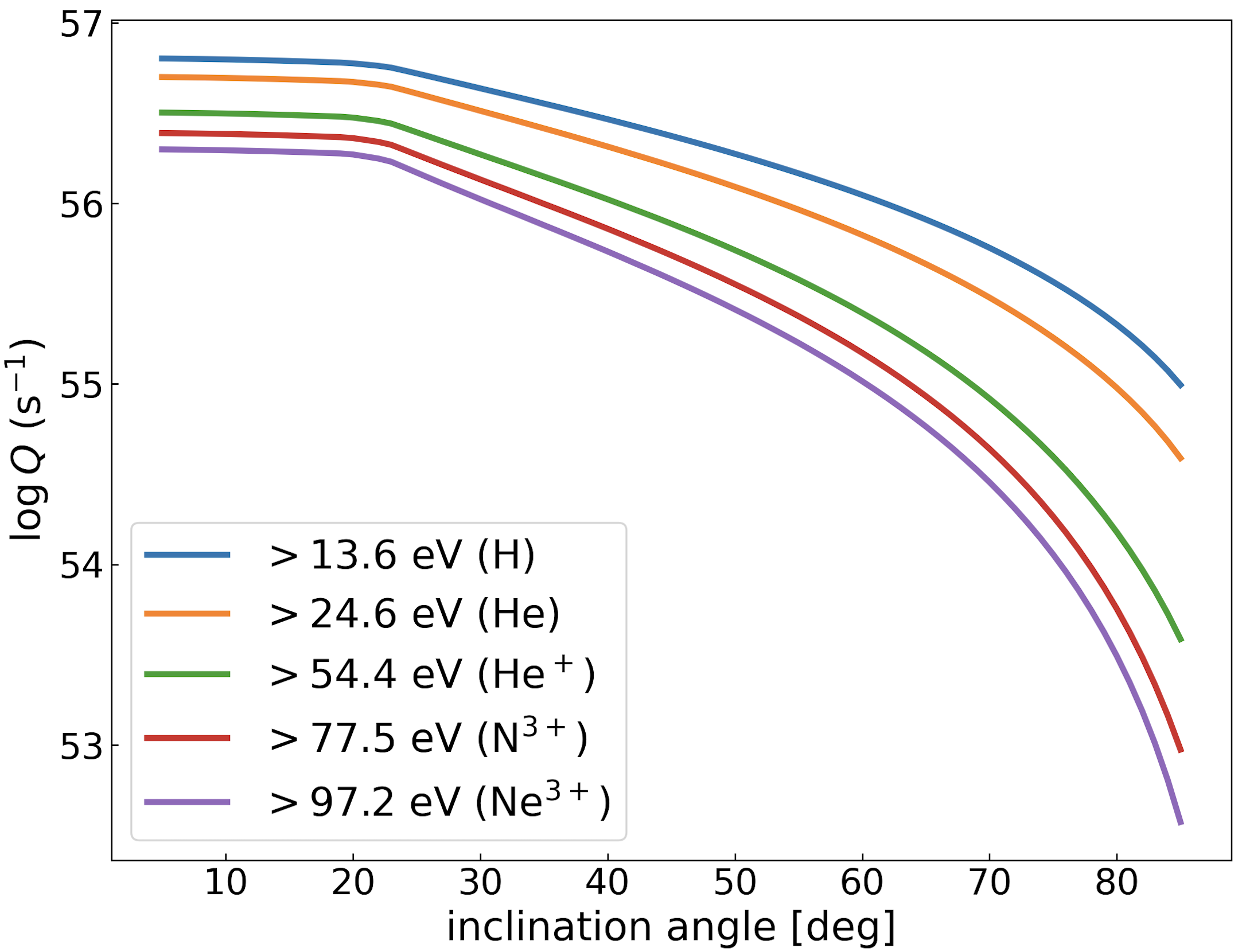}
\caption{Angle-dependent ionizing photon production rates, $Q(>E_{\rm th})$, for a super-Eddington accretion flow with $M_{\rm BH}=10^{7.5}\,M_\odot$ and $\dot m=32$ (Model A of \citealt{Madau2026}). The curves show the integrated photon rate (s$^{-1}$) above five ionization thresholds:
H{\sc i} (13.6 eV), He{\sc i} (24.6 eV), He{\sc ii} (54.4 eV), N{\sc iv} (77.5 eV), and Ne{\sc iv} (97.2 eV). The rates are strongly anisotropic, remaining near their peak values for $i\lesssim 25^\circ$ but declining steeply at high inclinations as self-shadowing obscures the hot inner funnel and the emergent spectrum becomes dominated by the cooler outer disk. From face-on to edge-on, the highest-threshold photon rates decrease by $\gta\,$2--3 dex.
}
\label{fig:hardness}
\end{figure}
 
\subsection{Anisotropic intrinsically blue SEDs}

Following \citet{Madau2026}, we compute inclination-dependent, scattering-modified blackbody SEDs for a fiducial $M_{\rm BH}=10^{7.5}\,M_\odot$ accretor at $\dot m\simeq 32$. The key outcome is a strongly anisotropic ionizing continuum: the EUV/soft-X photon output is highest for near-polar sightlines and is rapidly suppressed toward edge-on views by geometric foreshortening and self-shadowing, while the UV-optical continuum varies much more weakly with inclination. In the remainder of the paper we therefore characterize orientation-dependent ``hardness'' using photon-rate ratios above selected ionization thresholds (Fig.~\ref{fig:hardness}), which directly track the supply of H- and He-ionizing photons and the availability of the hardest EUV photons relevant for high-ionization species. 

We note that these models can produce a very blue rest-UV continuum, comparable to 
-- and in some cases steeper than -- the standard thin-disk asymptotic slope $f_\lambda\propto \lambda^{-7/3}$ ($\beta=-2.33$). In the $(\beta_{\rm UV},\beta_{\rm opt})$ plane, several JWST LBDs extend to very blue continua, approaching the slopes expected for an unreddened accretion disk \citep{Hainline2025,Brazzini2026}. Interestingly, fitting the NIRSpec optical continuum of the Red Rosetta Stone GN--28074 with a dust-absorbed power-law, \citet{Brazzini2026} infer an intrinsic slope $\beta=-2.49\pm 0.15$, i.e. close to accretion disk expectations. 

\subsection{Accretion flow geometry and the BLR}

We adopt a simple equatorial BLR geometry motivated by reverberation-mapping constraints \citep[e.g.,][]{Du2025,Nunez2014} and microlensing distortions of broad-line profiles \citep[e.g.,][]{Hutsemekers2023,Savic2024,Gaskell2009}. In the present implementation, the BLR is represented as a clumpy, equatorially concentrated distribution whose line-of-sight obscuration is described probabilistically by an inclination-dependent escape probability $P_{\rm esc}(i)$ (see below). 

\begin{figure}[!htb]
\centering
\includegraphics[width=\hsize,trim= 0 0 0 4mm,clip]{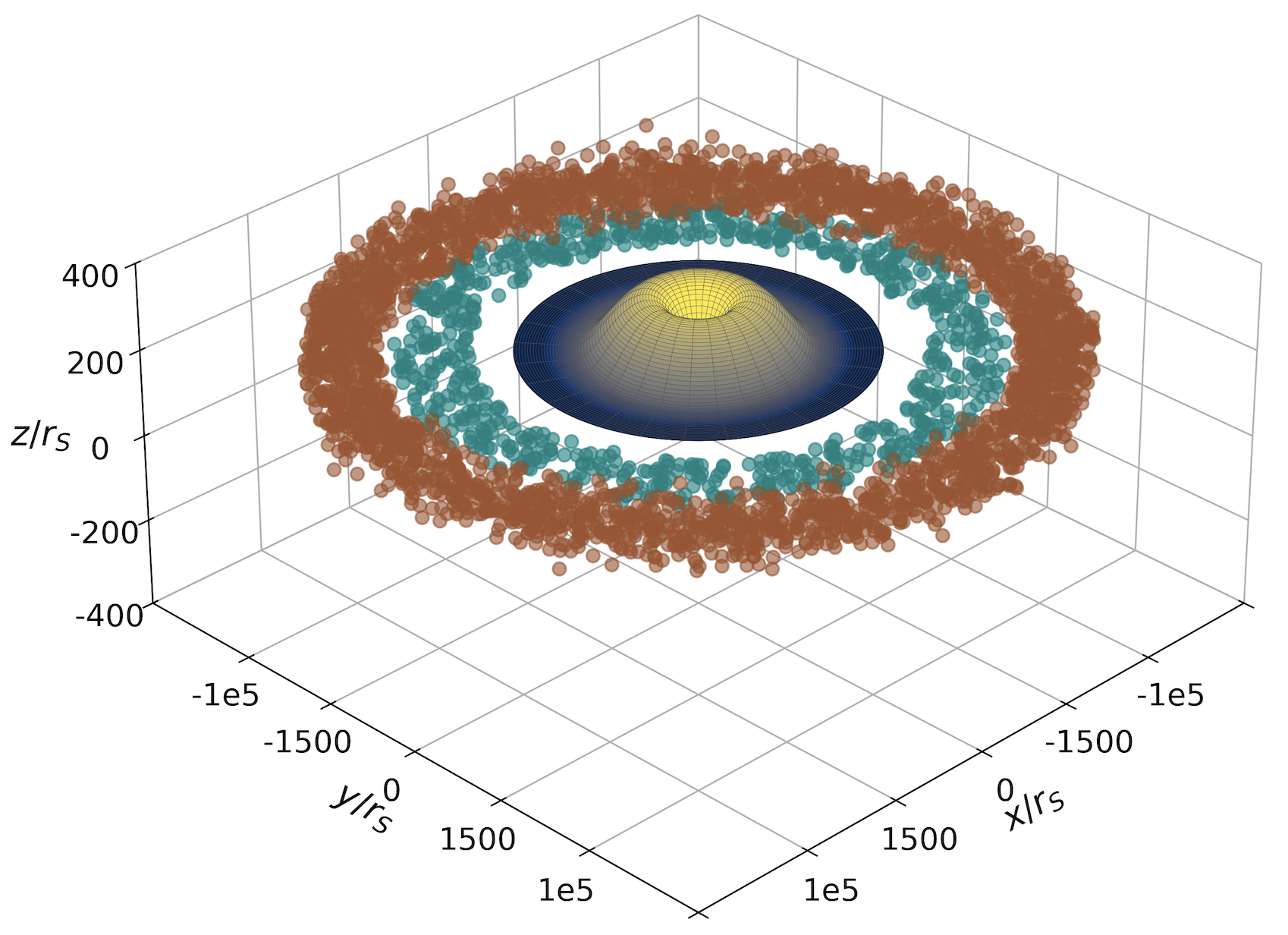}
\caption{Three-dimensional toroidal geometry of a super-Eddington BLAGN accreting at $\dot m=32$, viewed at an inclination angle of $i=60^\circ$. The spatial coordinates $x, y$, and $z$ are expressed in units of Schwarzschild radii $r_S$. The thick torus extends out to $\simeq 1000\,r_S$, where it is matched to an outer geometrically thin disk (only shown to $r=1500\,r_S$ to illustrate the geometric transition). Brighter (yellow) colors indicate higher surface brightness, while darker (blue) colors indicate lower values. The teal spheres provide a schematic (not-to-scale) visualization of BLR clouds; physically, the BLR lies at much larger radii ($\sim 10^5\,r_S$), and its outer portion beyond the sublimation radius is expected to be dusty. The outer brown structure schematically represents a dusty circumnuclear obscurer (a torus-like structure) surrounding the BLR.}
\label{fig:disk+BLR}
\end{figure}

In super-Eddington funnel geometries, gas near the equatorial plane is illuminated by a softer, self-shielded ionizing continuum than that seen along polar lines of sight, naturally suppressing high-ionization emission while preserving strong Balmer recombination lines \citep{Madau2026}. Our baseline BLAGN geometry is depicted in Figure~\ref{fig:disk+BLR} and includes an inner, geometrically thick torus extending to $r_{\rm out}\simeq 1000\,r_{\rm S}$. The BLR is assumed to reside at much larger radii in an equatorial distribution, with its outer portion beyond the sublimation radius embedded in (or contiguous with) a dusty circumnuclear torus (see below). To extend the SED beyond the outer edge of the supercritical funnel, we match the thick-disk solution to a standard radiation-pressure dominated thin disk by setting a transition radius $r_t=0.93\,r_{\rm out}$ and enforcing continuity of the effective temperature and radiative flux at $r_t$; outside $r_t$ the thin-disk flux follows the standard Keplerian scaling $F(r)\propto r^{-3}$ and is intrinsically subdominant in the UV. We truncate the outer disk at $r=1500\,r_{\rm S}$ near the expected self-gravity radius.

\noindent

\section{Extreme H$\alpha$ EWs: blue SEDs vs. covering factor}

The extreme Balmer-line strengths of LRDs \citep{Hviding2025} motivate a direct test of whether their large H$\alpha$ EWs are compatible with standard BLR covering factors given the ionizing SEDs predicted by our super-Eddington models.

{Grids of photoionization models were generated with the C23 release of {\sc Cloudy} \citep{Chatzikos2023}, tailored to typical BLR conditions ($n_{\rm H}=10^{10}\,{\rm cm^{-3}}$, $N_{\rm H}=10^{23}\,{\rm cm^{-2}}$). We varied the ionization parameter $U$ and adopted a metallicity representative of galaxies at the redshifts where LRDs and LBDs are most commonly found, $Z=0.1\,Z_\odot$. 

We model the BLR as a clumpy, equator-concentrated cloud distribution by specifying the mean number of clouds intersected along a line of sight at inclination $i$ as

\begin{equation}
N(i)=N_0\,
\exp\left[-\frac{\cos^2 i}{2\sigma_{\rm BLR}^2}\right],
\label{eq:Nlos}
\end{equation}

(see, e.g., the formalism of \citealt{Nenkova2008}), where $\sigma_{\rm BLR}$ controls the angular thickness of the cloud distribution about the equatorial plane. The corresponding escape probability for direct disk photons is

\begin{equation}
P_{\rm esc}(i)=\exp\big[-N(i)\big].
\label{eq:Pof_i}
\end{equation}

We fix the normalization $N_0$ by requiring that the angle-averaged probability of intercepting at least one cloud equals the global covering factor,

\begin{equation}
C_{\rm BLR}=\int_{0}^{\pi/2}\Big[1-P_{\rm esc}(i)\Big]\sin i\,{\rm d}i,
\label{eq:Cblr_norm}
\end{equation}

where we have used symmetry about the mid-plane. In practice, for a given covering factor $C_{\rm BLR}$ and angular width $\sigma_{\rm BLR}$, Equation (\ref{eq:Cblr_norm}) is solved for $N_0$. For $\sigma_{\rm BLR}\lta 0.2$, the distribution is strongly equator-weighted, with a median intercepted inclination exceeding $80^\circ$.}

{For each observer inclination $i$, we compute the rest-frame EW of a line at wavelength $\lambda_0$ as

\begin{equation}
{\rm EW}(i)=\frac{C_{\rm BLR}\,\overline{F}_{\rm line}}{F_{\lambda,{\rm cont}}(\lambda_0;i)},
\label{eq:EW_def}
\end{equation}

where $\overline{F}_{\rm line}$ is the BLR-averaged broad-line flux per unit covering factor and is independent of the observer inclination. It is obtained by interpolating the {\sc Cloudy} line emissivities over a grid of angles ($65^\circ,70^\circ,75^\circ,80^\circ,85^\circ$) and averaging them over the adopted equatorially concentrated angular distribution of BLR clouds. Here the same angular variable $i$ is used to specify both the BLR cloud distribution and the observer inclination; however, the broad-line flux $\overline{F}_{\rm line}$ and diffuse continuum $\overline{F}_{\lambda,{\rm diff}}$ are averaged over the BLR angular distribution and are therefore independent of the particular observer's line of sight. Thus, the numerator reflects the global BLR geometry and covering factor rather than the observer's particular line of sight.}

{The continuum flux density per unit wavelength at the line, $F_{\lambda,{\rm cont}}(\lambda_0;i)$, is obtained from the observed continuum spectrum as

\begin{equation}
\begin{split}
F_{\lambda,{\rm cont}}(\lambda;i)=\;&
P_{\rm esc}(i)\,F_{\lambda,{\rm disk}}(\lambda;i) +\big[1-P_{\rm esc}(i)\big]\\
&\times F_{\lambda,{\rm tran}}(\lambda;i)
+C_{\rm BLR}\,\overline{F}_{\lambda,{\rm diff}}(\lambda),
\end{split}
\label{eq:Fcont_decomp}
\end{equation}

evaluated at $\lambda=\lambda_0$. Here $F_{\lambda,{\rm disk}}(\lambda;i)$ is the direct, inclination-dependent disk continuum from the accretion flow, $F_{\lambda,{\rm tran}}(\lambda;i)$ is the transmitted continuum along the observer's line of sight through BLR gas, and $\overline{F}_{\lambda,{\rm diff}}(\lambda)$ is the diffuse BLR continuum, treated as isotropic and therefore independent of observer inclination. The direct and transmitted nuclear continua are weighted by the line-of-sight escape probability $P_{\rm esc}(i)$, while $C_{\rm BLR}$ sets the amplitude of the diffuse BLR continuum.}

\begin{figure}[!htb]
\centering
\includegraphics[width=\hsize,trim=0 0 0 0mm,clip]{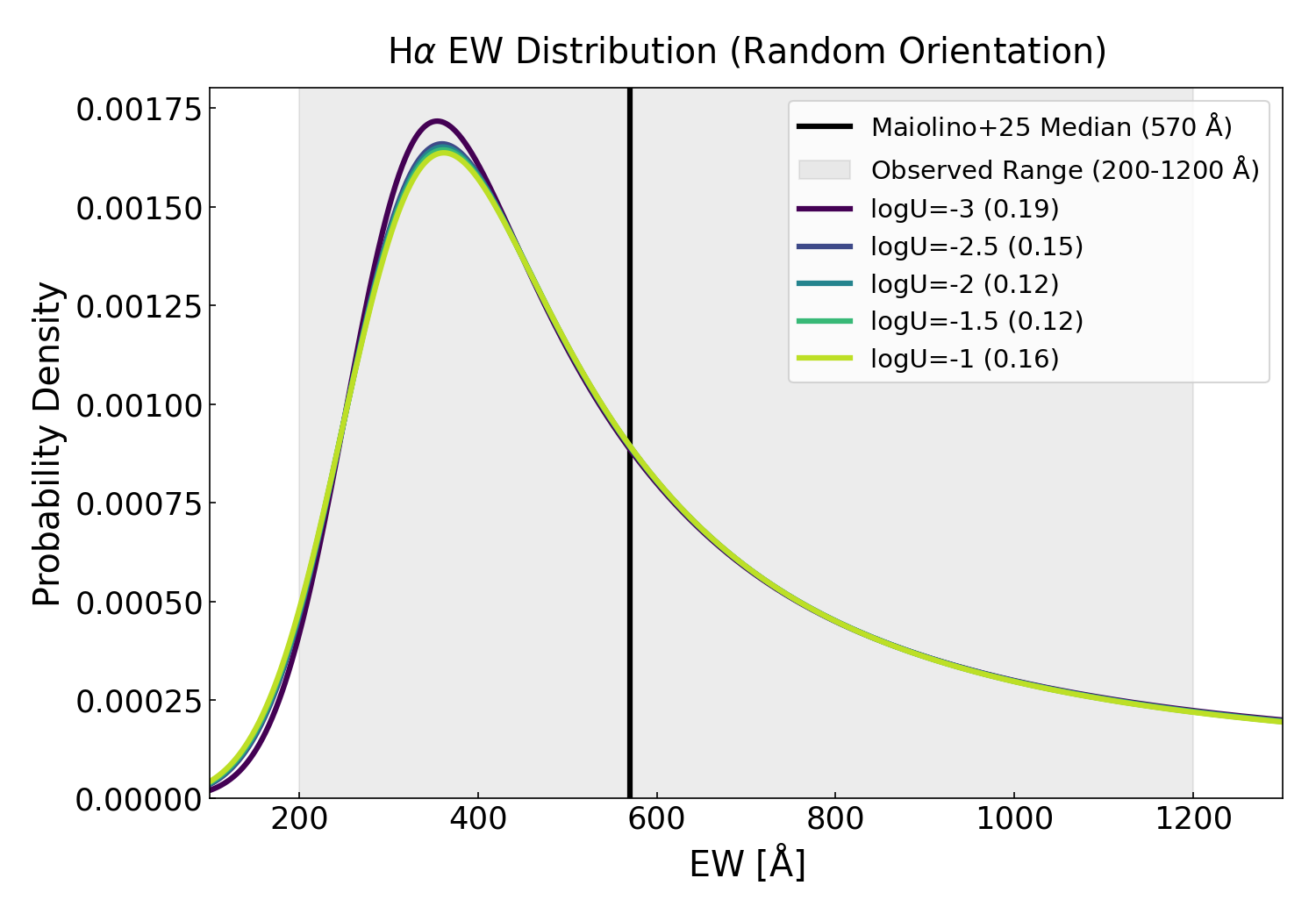}
\caption{Probability density functions (PDFs) of the broad H$\alpha$ EW predicted by our fiducial super-Eddington accretor with $M_{\rm BH}=10^{7.5}\,M_\odot$ and $\dot m=32$. The distributions are derived by weighting the inclination-dependent EWs by the geometric probability $p(i)=\sin i$ ($0\le i\le \pi/2$), {assuming random observer orientations and a fiducial BLR angular thickness of $\sigma_{\rm BLR}=0.2$ (the Gaussian scale height of the cloud distribution about the equatorial plane)}. Different curves correspond to different ionization parameters $\log U$, as indicated in the legend. For each model, the global BLR covering factor $C_{\rm BLR}$ is scaled such that the probability-weighted median of the theoretical distribution matches the observed median EW of $570$\,\AA\ from the high-$z$ BLAGN sample of \citet{Maiolino2025} (vertical black line). The resulting required covering factors are listed in parentheses in the legend and are relatively modest ($C_{\rm BLR}=0.10$--0.20). The gray shaded region denotes the full range of EWs observed in the JADES sample ($200$--$1200$\,\AA). 
}
\label{fig:PDF}
\end{figure}

Figure~\ref{fig:PDF} displays the resulting broad H$\alpha$ EW probability density functions (PDFs), computed assuming random observer orientations weighted by the random-orientation distribution $p(i)=\sin i$. The shaded region indicates the full range of EWs observed in the high-$z$ sample of \citet{Maiolino2025} ($200 \lesssim {\rm EW} \lesssim 1200$ \AA). While the observed histogram is a population statistic that also reflects intrinsic scatter in black hole mass, accretion rate, and BLR gas properties, our model curves isolate the variance due to orientation alone and are therefore narrower. Nevertheless, the shift of the probability-weighted peak EW with ionization parameter provides a useful handle on the overall scaling needed to reproduce typical line strengths.
In the most edge-on tail of the orientation distribution, the model predicts very large EWs because the observed optical continuum decreases approximately geometrically, $\propto \cos i$, while the broad-line luminosity is assumed to be isotropic. In practice, once the direct nuclear continuum becomes extremely faint -- in our models this occurs at $i \gtrsim 85^\circ$ -- the observed rest-optical continuum will not continue to dim indefinitely. A low-level continuum floor from host starlight and/or scattered AGN light is expected to dominate over the highly foreshortened direct disk emission at these inclinations, effectively truncating the high-EW tail. We therefore impose a conservative cap ${\rm EW}(\mathrm{H}\alpha)\le 2000$\,\AA\ when constructing the PDFs. 

Matching the probability-weighted median of the models to the observed median implies global covering factors of $C_{\rm BLR}\simeq 0.1$--0.2. These values are consistent with standard estimates for low-$z$ Type~1 AGNs \citep[e.g.,][]{Peterson2006,Pandey2023}. In this framework, $C_{\rm BLR}$ is simply the sky fraction covered by BLR clouds as seen from the central engine, i.e. the angle-averaged probability that an emitted photon encounters at least one cloud. For a clumpy, non-axisymmetric distribution, $C_{\rm BLR}$ is an average over azimuth and polar angle of the cloud covering probability, and does not imply unity covering along any specific subset of sightlines.

The EW distribution reported by \citet{Maiolino2025} combines objects spanning the emerging LBD/LRD taxonomy, including JADES sources as well as systems drawn from the literature. In our framework, LBDs and LRDs occupy different parts of this distribution: less-reddened LBDs should preferentially exhibit lower EWs, whereas obscured, high-inclination LRDs should populate the high-EW tail. This prediction is directly testable by measuring EW distributions separately for LBD and LRD subsamples. 

\begin{figure}[!htb]
\centering
\includegraphics[width=\hsize,trim=0 0 0 0,clip]{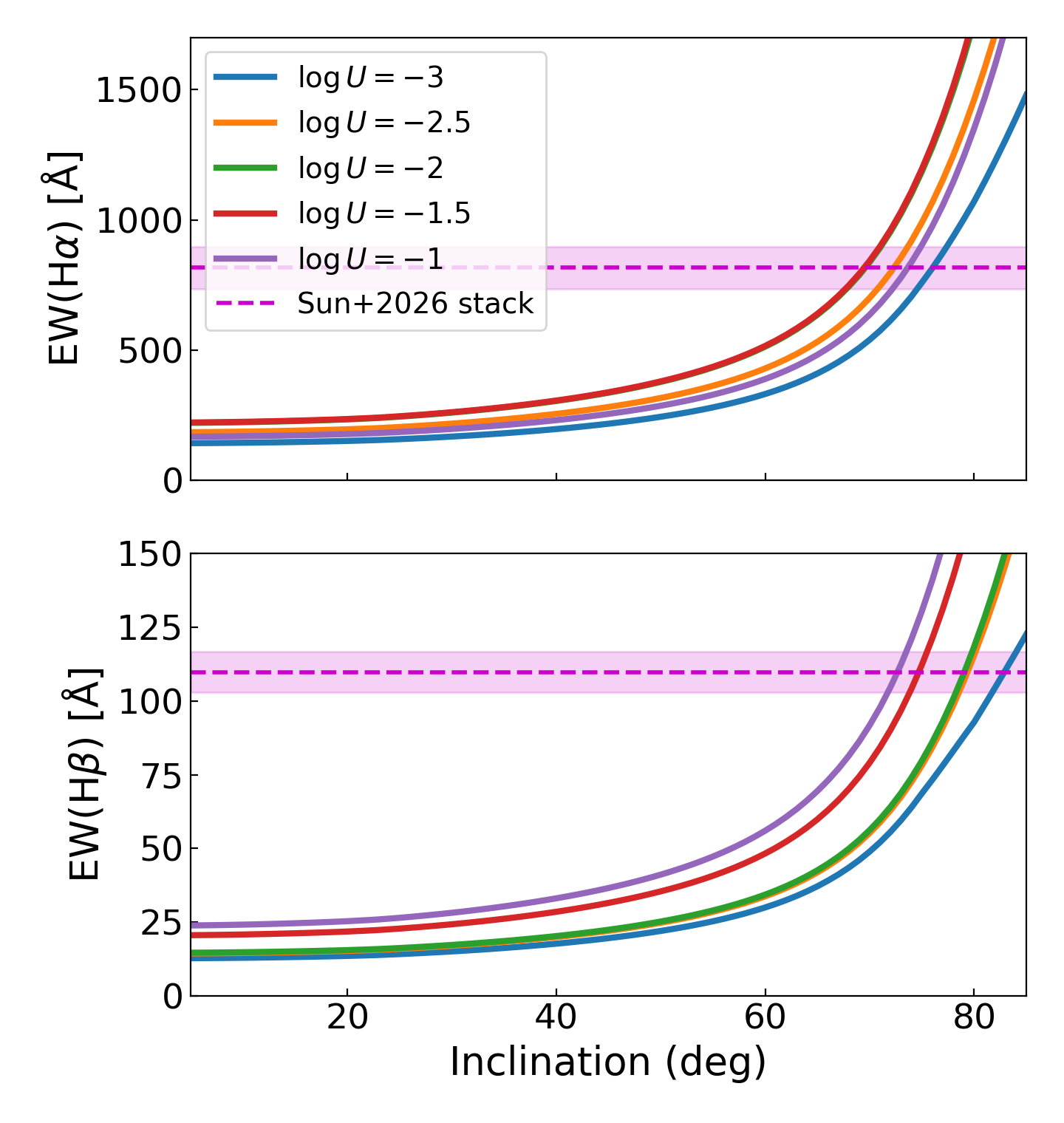}
\caption{Predicted inclination dependence of the Balmer EWs. The top and bottom panels show ${\rm EW}(\mathrm{H}\alpha)$ and ${\rm EW}(\mathrm{H}\beta)$, respectively, as functions of observer inclination for our fiducial super-Eddington model with $M_{\rm BH}=10^{7.5}\,M_\odot$ and $\dot m=32$.
Curves correspond to different ionization parameters $\log U$ (legend), computed for a fixed BLR covering factor $C_{\rm BLR}=0.1$. The magenta dashed lines and shaded bands indicate the total (broad+narrow) EWs measured from the median-stacked LRD spectrum of \citet{Sun2026}, ${\rm EW}(\mathrm{H}\alpha)=817.2^{+80.9}_{-82.7}\,\text{\AA}$ and ${\rm EW}(\mathrm{H}\beta)=109.8^{+6.8}_{-6.9}\,\text{\AA}$; at the PRISM resolution these values cannot in general be decomposed into broad and narrow components. In our model, matching the large Balmer EWs favors highly inclined sightlines, typically $i\gtrsim 70^\circ$ depending weakly on $\log U$.
}
\label{fig:Ha_Hb.png}
\end{figure}

{A recent median-stacked LRD spectrum from \citet{Sun2026} yields total (broad+narrow) equivalent widths ${\rm EW}(\mathrm{H}\alpha)=817.2^{+80.9}_{-82.7}\,$\AA\ and ${\rm EW}(\mathrm{H}\beta)=109.8^{+6.8}_{-6.9}\,$\AA. We note that this LRD-specific value already lies well above the combined LRD+LBD population median of $570\,$\AA\ reported by \citet{Maiolino2025}, hinting at the LRD/LBD EW ordering predicted by our model even before a direct, broad-line comparison of the two subpopulations was available. In our inclination-only model, such large Balmer EWs are naturally attained only for highly inclined sightlines. Figure~\ref{fig:Ha_Hb.png} shows that matching the Sun et al.\ stack requires inclinations $i\gtrsim 70^\circ$ (depending weakly on $\log U$), with lower-inclination views underpredicting both ${\rm EW}(\mathrm{H}\alpha)$ and ${\rm EW}(\mathrm{H}\beta)$. This may support the interpretation that color-selected LRDs correspond to the high-inclination tail of the same underlying super-Eddington BLAGN population whose less-reddened, more face-on analogs appear as LBDs. We caution, however, that these stacked measurements correspond to total (broad+narrow) EWs at the PRISM resolution, where a robust decomposition into broad and narrow components is not generally possible; the broad-line EWs relevant for our BLR modeling may therefore be somewhat smaller \citep{Geris2026}.
}

Our results contrast with \citet{Yan2025}, who argue that the observed broad H$\alpha$ strengths require near-unity covering factors and therefore invoke an enshrouded geometry with nearly $4\pi$ coverage by dense gas ($n_{\rm H}=10^8$--$10^{10}\,{\rm cm^{-3}}$). In our super-Eddington models, large Balmer EWs can be produced with modest $C_{\rm BLR}$ because the BLR-illuminating continuum has a higher ionizing-to-optical photon budget than standard quasar composites, and the inclination dependence is driven by anisotropic continuum dilution at fixed line emission. For more face-on sightlines, the observer sees the boosted funnel continuum, whereas the BLR does not intercept a proportionally larger ionizing flux, so the lines are diluted and the EWs fall to values typical of Type~1 AGNs. Thus, the most extreme LRD H$\alpha$ EWs require high inclinations, not unusually large covering factors.

{While this paper was under review, \citet{Geris2026} presented a new analysis of the H$\alpha$ EW distributions of LBD and LRD subsamples. They found that LRDs exhibit systematically larger H$\alpha$ EWs and larger H$\alpha$/H$\beta$ ratios than LBDs, providing the first direct, same-sample confirmation of the predicted trend. We discuss these results in Section~4.}

\noindent
\begin{figure}[b]
\centering
\includegraphics[width=\hsize, trim=0 0 0 0 ,clip]{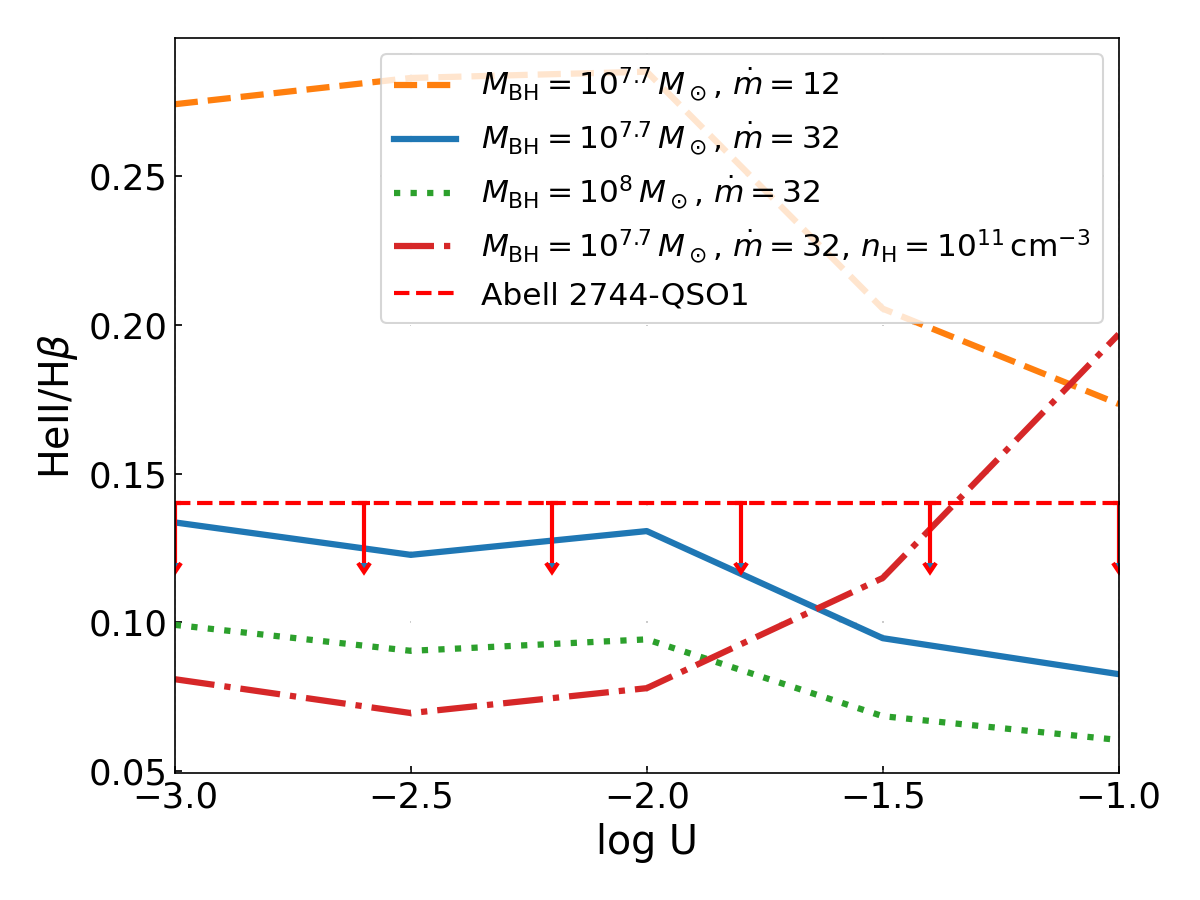}
\caption{
Predicted He{\sc ii} $\lambda4686/\mathrm{H}\beta$ as a function of $\log U$ for super-Eddington SEDs of different $M_{\rm BH}$ and $\dot m$ (see legend). Unless otherwise noted, $n_{\rm H}=10^{10}\,{\rm cm^{-3}}$; the single alternative-density case is labeled in the legend. The dashed line shows the upper limit for the LRD Abell~2744--QSO1 from JWST spectroscopy \citep{Furtak2024,Ji2025}. All models assume $Z=0.1\,Z_\odot$ and $N_{\rm H}=10^{23}\,{\rm cm^{-2}}$.
}
\label{fig:HeII_Hbeta}
\end{figure}

\subsection{The missing hard photons problem of LRDs}

One proposed explanation for the extreme X-ray weakness of JWST-discovered BLAGNs invokes heavy (Compton-thick) absorption by dense gas along the line of sight \citep[e.g.,][]{Maiolino2025,Ji2025}. Columns as large as $N_{\rm H}\gtrsim 10^{25}\,{\rm cm^{-2}}$ must have a very low dust-to-gas ratio to allow for the detection of broad lines; BLR clouds could, in principle, satisfy this requirement. In LRDs, this interpretation has been linked to reports of deep Balmer absorption and non-stellar Balmer breaks, which imply high-density gas along the line of sight \citep{Juod2024JADES,Deugenio2026_restframeabs,Ji2025,Naidu2025}. However, the same explanation is difficult to extend to LBDs, which are also X-ray undetected even in stacking analyses, yet show no evidence for Balmer absorption or other signatures of dense intervening gas \citep{Maiolino2025,Brazzini2026}.

A related puzzle is that, while JWST spectra of LRDs and LBDs show strong broad Balmer emission, high-ionization lines such as He{\sc ii}, C{\sc iv}, and N{\sc v} are often weak or undetected, pointing to an ionizing continuum incident on the line-emitting gas that differs from a standard quasar power-law SED \citep{Zucchi2026}. One proposed explanation is the ``gas-enshrouded black hole'' scenario \citep[e.g.,][]{Kido2025,Naidu2025}, in which a compact, extremely dense H{\sc i} envelope absorbs EUV and X-ray photons and reprocesses them into a softer continuum. This picture faces basic geometric and radiative-transfer difficulties: if the cocoon lies inside the BLR, it would severely limit the escape of ionizing photons, implying that the BLR must be ionized externally; if it lies outside the BLR, the broad lines should be strongly suppressed or hidden, since the putative atmosphere is assumed to be optically thick, thermalized, and to emit approximately as a blackbody with $T\sim 5000\,{\rm K}$.

Alternatively, in super--Eddington funnel geometries, the X-ray weakness is naturally explained by a Compton-cooled corona and, additionally, by orientation-dependent spectral shaping \citep{Madau2026}. In these models the EUV band ($600$--$900\,$\AA) remains blue and only moderately inclination-dependent, with $\beta_{\rm EUV}\simeq -2.1$ for face-on views and flattening to $\beta_{\rm EUV}\simeq -1.5$ near the equatorial plane, whereas the harder XUV band ($200$--$600\,$\AA), which controls the production of high-ionization species, is strongly suppressed in equatorial directions and its slope changes from $\beta_{\rm XUV}\simeq -1.9$ face-on to $\beta_{\rm XUV}\simeq -0.14$ near the equatorial plane. Around the He$^{+}$ edge ($54.4$ eV), this strong XUV softening naturally reduces He{\sc ii} (and related) emission while preserving strong Balmer lines.

Figure~\ref{fig:HeII_Hbeta} compares our \textsc{Cloudy} predictions for He{\sc ii}$\,\lambda4686/\mathrm{H}\beta$ to the observational upper limit for the LRD Abell~2744--QSO1. The model curves show the predicted ratio as a function of ionization parameter for super-Eddington SEDs with different black hole masses and accretion rates (see legend; the dot-dashed curve shows the single higher-density case). The hardest track ($M_{\rm BH}=10^{7.7}\,M_\odot$, $\dot m=12$) exceeds the upper limit across the full $\log U$ range, whereas the softer SEDs at higher $\dot m$ and/or larger $M_{\rm BH}$ satisfy the constraint. The black hole mass directly measured in Abell~2744--QSO1 by \citet{Juodzbalis2025_QSO1} is $\log (M_{\rm BH}/M_\odot)\simeq 7.7$, consistent with single-epoch measurements \citep{Furtak2024,Ji2025,Deugenio2025}. The relatively low $\lambda_{\rm Edd}$ reported in previous work likely reflects standard bolometric corrections tied to broad H$\alpha$ and local quasar SED templates; dust attenuation and SED anisotropy can bias such inferences and allow for substantially higher intrinsic $\lambda_{\rm Edd}$ in our framework (see below).

\subsection{Dust reddening,V-shaped SEDs, and Balmer decrements}

In our orientation-based picture linking LBDs and LRDs, the two populations correspond to the same underlying BLAGN class viewed at different inclinations: LBDs are preferentially seen intermediate-lower inclinations, while LRDs are those BLAGNs observed closer to the equatorial plane. Along these high-inclination sightlines, substantial dust attenuation is required to transform an intrinsically blue, 
super-Eddington AGN SED into the observed red rest-optical continuum. A number of recent analyses of LRDs have invoked extinctions of $A_V\sim$ a few magnitudes \citep[e.g.,][]{Killi2024,Brooks2025,Li2025,Schindler2025,Akins2025,Ji2025,Brazzini2026}. In this framework, the reddening is naturally attributed to an equatorial dusty torus located on the outskirts of the BLR or immediately beyond it, where the gas lies outside the dust sublimation radius and can sustain a significant dust column. Such a configuration provides a direct physical connection between the persistence of broad emission lines and the strong continuum reddening: for foreground dusty clouds, both the BLR lines and the local continuum at a given wavelength are attenuated by the same factor (so the EW is preserved), while the overall SED becomes strongly reddened and bluer broad-line and continuum features are preferentially suppressed.

The dusty obscurer is modeled with the same clumpy formalism as the BLR (Eqs.~1--3), but with its own parameters $(C_{\rm dust},\sigma_{\rm dust})$. Physically, a broader (``flared'') angular distribution for the dusty component is expected because the dust-bearing obscurer resides at larger radii (beyond the dust sublimation front), where vertical support from radiation pressure on dust and/or disk winds can produce a larger scale height than in the dust-free BLR \citep[e.g.,][]{Elitzur2006,Nenkova2008,Elitzur2008,Wada2012}. Concretely, we take

\begin{equation}
N_{\rm dust}(i)=N_{0,d}\,
\exp\!\left[-\frac{\cos^2 i}{2\sigma_{\rm dust}^2}\right],
\label{eq:Nlos_dust}
\end{equation}

and

\begin{equation}
P_{{\rm esc},d}(i)=\exp\!\big[-N_{\rm dust}(i)\big],
\label{eq:Pdust}
\end{equation}

and determine $N_{0,d}$ by requiring that the solid-angle averaged probability of encountering at least one dusty cloud equals the global dust covering factor,

\begin{equation}
C_{\rm dust}=\int_{0}^{\pi/2}\big[1-P_{{\rm esc},d}(i)\big]\sin i\,{\rm d}i.
\label{eq:Cdust}
\end{equation}

{\citet{MadauMaiolinoLF} have recently shown that the proposed framework reproduces both the observed relative abundance of LRDs and LBDs and their luminosity distribution using only a modest luminosity-dependent dust covering factor, with an average value of order $\langle C_{\rm dust}\rangle\simeq 0.2$, together with characteristic cloud extinctions of a few magnitudes and a dusty distribution that is more flared than the BLR. 

In this scheme, $C_{\rm dust}$ determines the overall fraction of dust-intersecting sightlines, while their inclination distribution is given by the solid-angle weighted conditional probability

\begin{equation}
p(i\,|\,d) \propto \big[1-P_{{\rm esc},d}(i)\big]\sin i,
\label{eq:pidust}
\end{equation}

that is, the random orientation distribution ($\propto\sin i$) reweighted by the probability of intersecting at least one dusty cloud. Throughout this section, because our SED comparisons are conditioned on LRD selection, we fit for the extinction of a single dusty cloud, $A_V$, and compute the effective attenuation as the conditional, Poisson-averaged transmission

\begin{equation}
T_{\rm cond}(\lambda;i)=\frac{\exp\!\big[-N_{\rm dust}(i)\,(1-T_c(\lambda))\big]
-P_{{\rm esc},d}(i)}{1-P_{{\rm esc},d}(i)},
\label{eq:Tcond}
\end{equation}

where $T_c(\lambda)$ is the transmission of a single cloud of extinction $A_V$, evaluated at the representative inclination $i$ derived above, rather than additionally sampling a stochastic extinction realization for each sightline.

The intrinsic SED (continuum plus emission lines) at inclination $i$ is mapped to the observed one via

\begin{equation}
\begin{split}
\qquad\big[F_{\lambda,{\rm cont}}(\lambda;i)& +\overline{F}_{\lambda,{\rm line}}(\lambda)\big]
\;\longrightarrow\;\\
\big[F_{\lambda,{\rm cont}}(\lambda;i)& +\overline{F}_{\lambda,{\rm line}}(\lambda)\big]\,
T_{\rm cond}(\lambda;i),
\end{split}
\label{eq:dust_att}
\end{equation}

where $F_{\lambda,{\rm cont}}(\lambda;i)$ is given by
Equation~(\ref{eq:Fcont_decomp}) and already includes the diffuse BLR
continuum contribution $C_{\rm BLR}\,\overline{F}_{\lambda,{\rm diff}}(\lambda)$,
and $T_{\rm cond}(\lambda;i)$ is given by Equation~(\ref{eq:Tcond}),
evaluated at the fitted per-cloud extinction $A_V$ and the representative inclination $i$ derived above.}

{We fix the central-engine and BLR parameters ($M_{\rm BH}$, $\dot{m}$, $C_{\rm BLR}$, $\sigma_{\rm BLR}$), the dusty-cloud angular distribution width and covering factor ($\sigma_{\rm dust}, C_{\rm dust}$), and the ionization parameter $\log U$ at their fiducial values (given in Table~\ref{tab:dustfit}), and perform a least-squares fit of the resulting model spectrum to the composite LRD photometry of \citet{Delvecchio2025}, varying
only the three free parameters of the dusty cloud -- $A_V$, $R_V$, and the UV-bump strength $B$ in the flexible \citet{Cardelli1989}+
\citet{Conroy2010} parameterization. The observer inclination, $i=70^\circ$, is the conditional median inclination of dust-intersecting (LRD-type) sightlines implied by $\sigma_{\rm dust}=0.5$, and we therefore treat it as a derived quantity rather than an additional fixed parameter. 
As an additional, largely independent check on this choice we note that, for our adopted $C_{\rm BLR}$ and $\sigma_{\rm BLR}$, the model EW(H$\alpha$) at $i=70^\circ$ ($\simeq830$\,\AA) closely matches the LRD-stack value of $817.2^{+80.9}_{-82.7}$\,\AA\ measured by \citet{Sun2026}; since EW(H$\alpha$) is highly sensitive to inclination in our model but invariant under foreground dust reddening, this provides supporting evidence for $i\simeq 70^\circ$ that is complementary to, and not directly degenerate with, the photometric dust fit below.}

{Model and data are normalized to each other at a fixed anchor wavelength of
$7800$\,\AA\ (rest frame); this normalization is not a free parameter of the fit. The fit uses the $N=10$ photometric points spanning $1250\,\mbox{\AA}$--$1.4\,\mu$m (3 free parameters, $\nu=7$). Table~\ref{tab:dustfit} summarizes all parameters entering this calculation, distinguishing those fixed at the fiducial values adopted throughout this paper and the three dusty cloud
parameters constrained here. The best fit gives $A_V=3.1$, $R_V=4.1$, and
$B=0.2$ -- i.e.\ a relatively gray extinction curve with a strongly suppressed 2175\,\AA\ feature. Such a curve is physically plausible if the dust in these high-$z$ systems is metal-poor and/or embedded in dense, processed environments with
grain properties that differ from the local Milky Way ISM (see also
\citealt{Killi2024,Li2025}). The resulting curve remains steeper in the far-UV ($\lambda<3000$\,\AA) than the notably flat AGN extinction law of
\citet{Gaskell2004}.}

\begin{table}
\caption{Parameters entering the AGN+BLR+dust model fit to the composite LRD photometry of \citet{Delvecchio2025} (Figure~\ref{fig:dustSED}).}
\label{tab:dustfit}
\centering
\begin{tabular}{lcl}
\hline\hline
Parameter & Value & Comment \\
\hline
\multicolumn{3}{l}{Fixed} \\
$M_{\rm BH}$        & $10^{7.5}\,M_\odot$         & fiducial \\
$\dot{m}$           & $32$                        & fiducial \\
$\log U$            & $-1.5$                      & fiducial \\
$n_{\rm H}$         & $10^{10}\,{\rm cm}^{-3}$   & fiducial \\
$N_{\rm H}$         & $10^{23}\,{\rm cm}^{-2}$   & fiducial \\
$C_{\rm BLR}$       & $0.10$                      & fiducial \\
$\sigma_{\rm BLR}$  & $0.2$                       & fiducial \\
$\sigma_{\rm dust}$ & $0.5$                       & fiducial \\
$C_{\rm dust}$      & $0.2$                       & fiducial \\
\hline
\multicolumn{3}{l}{Derived} \\
$i$                 & $70^\circ$                  & median \\
\hline
\multicolumn{3}{l}{Free} \\
$A_V$               & $3.1$                       & best fit \\
$R_V$               & $4.1$                       & best fit \\
$B$                 & $0.2$                       & best fit \\
\hline
\multicolumn{3}{l}{Flux normalization} \\
Anchor wavelength & $7800$\,\AA\ (rest) & fixed \\
\hline
\end{tabular}
\end{table}

\begin{figure}[!ht]
\centering
\includegraphics[width=\hsize,trim=0 0 0 0,clip]{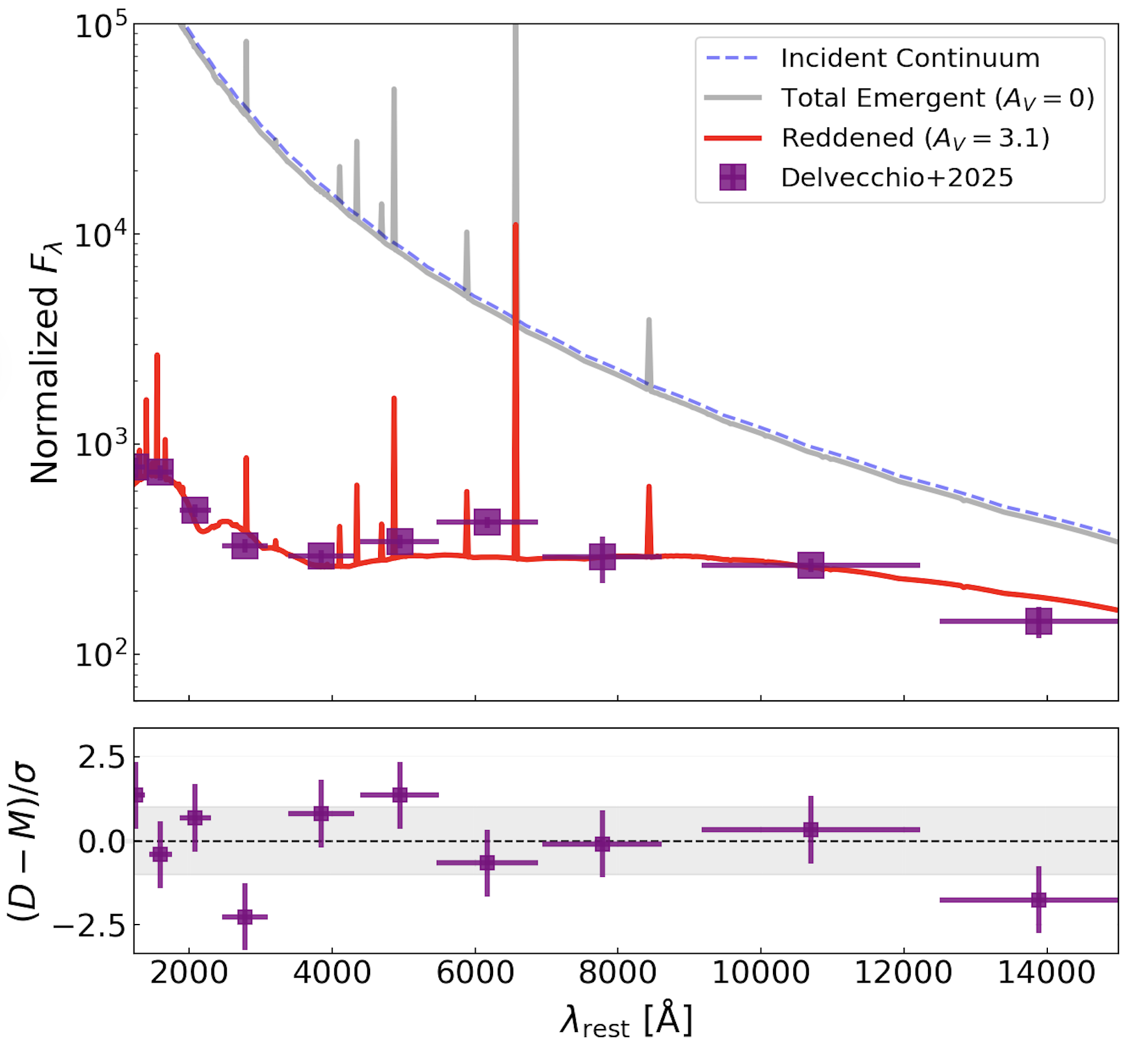}
\caption{Comparison of the stacked photometry of LRDs from \citet{Delvecchio2025} (purple squares) with our best-fit dust-attenuated accretion-disk model; the full suite of model parameters is given in Table~\ref{tab:dustfit}. Fluxes are normalized at 7800\,\AA. The gray line shows the total emergent spectrum (continuum plus lines) after BLR processing (computed with \textsc{Cloudy}; no dust attenuation applied), shown for an observer at the conditional median inclination $i=70^\circ$ of dust-intersecting sightlines for our adopted $\sigma_{\rm dust}$. The red line shows the spectrum after attenuation by foreground dusty clouds using a modified \citet{Cardelli1989} law, with $A_V$, $R_V$, and the bump strength $B$ obtained by jointly fitting the model to the $1250\,\mbox{\AA}$--$1.4\,\mu$m photometry (best fit: $A_V=3.1$, $R_V=4.1$, $B=0.2$; $\chi^2_\nu=1.9$ for $\nu=7$). The lower panel shows the corresponding residuals, $(D-M)/\sigma$, at each photometric point.}
\label{fig:dustSED}
\end{figure}

{Figure~\ref{fig:dustSED} compares the best-fit SED of our fiducial
super-Eddington model to the composite UV--near-IR SED of a large, homogeneously selected sample of LRDs from multiple JWST Legacy fields with median redshift $\langle z\rangle\simeq6.2$ \citep{Delvecchio2025}. The resulting V-shaped SED resembles the observed optical--UV spectrum of LRDs, with no evident need for a significant additional stellar contribution or AGN scattered light.\footnote{We suppress the intrinsic Ly$\alpha$ and He{\sc i}~$\lambda10830$ emission components in our model, motivated by the very large line optical depths expected in heavily obscured environments: Ly$\alpha$ photons undergo resonant trapping, while He{\sc i}~$\lambda10830$ can become optically thick due to the metastable $2^{3}S$ lower level. In both cases, repeated scatterings increase the effective path length and enhance the probability of absorption by dust, plausibly reducing the emergent line flux
in LRDs.}
We cannot exclude a contribution to the UV emission from star formation (especially if it is spatially extended and/or offset from the nucleus), but our fit does not require such a component, and the bulk of the UV light may still be attributed to transmitted AGN radiation. In cases where the observed UV emission is instead dominated by star formation, this can be accommodated by adopting a steeper extinction curve, which provides additional suppression of the AGN continuum in the UV (see, e.g., the case of The Cliff discussed later in \S~\ref{sec:Cliff}). Over the fitted range, the model reproduces the data with a mean model-to-data flux ratio of 1.02 and an rms scatter of 10\%, comparable to the level of intrinsic object-to-object variance expected in the parent LRD population and to bandpass-averaging effects on strong spectral features. Most of the ten fitted photometric points agree with the best-fit model to within a few percent, but three points drive most of the residual $\chi^2$ -- a 15\% model deficit at the bluest point ($\sim$1250\,\AA, coincident with the Ly$\alpha$ region), a 16\% model excess near
$2700$--$2800$\,\AA\ (coincident with the wavelength of population-averaged Mg\,{\sc ii}\,$\lambda2798$ BLR emission entering the corresponding photometric bandpass), and a 25\% model excess near $1.39\,\mu$m, which more likely reflects residual continuum-shape mismatch or intrinsic scatter in the stacked photometry.}

{With $\log U=-1.5$, $n_{\rm H}=10^{10}\,{\rm cm}^{-3}$, and
$N_{\rm H}=10^{23}\,{\rm cm}^{-2}$, the processed continuum can exhibit a
sizeable Balmer discontinuity, with a clear contribution from Balmer-edge
(continuum bound-free) absorption imprinted by the BLR gas on the
transmitted component. This feature is diluted in the composite stack by
inclination-dependent mixing with the direct disk continuum, and a
pronounced break is instead expected only for the most strongly attenuated, near-equatorial sightlines ($P_{\rm esc}(i)\ll1$), for which the transmitted component alone dominates the emergent continuum (see below). We find that the diffuse nebular continuum, $\overline{F}_{\lambda, {\rm diff}}$ in Equation (\ref{eq:Fcont_decomp}), remains subdominant to the transmitted component  at the covering factors considered here ($C_{\rm BLR}\lesssim 0.1$).
Once again, this applies to the average spectrum of the whole population; we cannot exclude that a few individual sources with larger BLR covering factor might have a larger nebular contribution, as indeed inferred from the Paschen jump in a few LRDs \citep{Sneppen2026Paj} and as inferred for Seyfert galaxies in the past \citep[e.g.][]{Korista2019,Netzer2022}.
}

\begin{figure}[!ht]
\centering
\includegraphics[width=0.95\hsize,trim=0 0 0 0,clip]{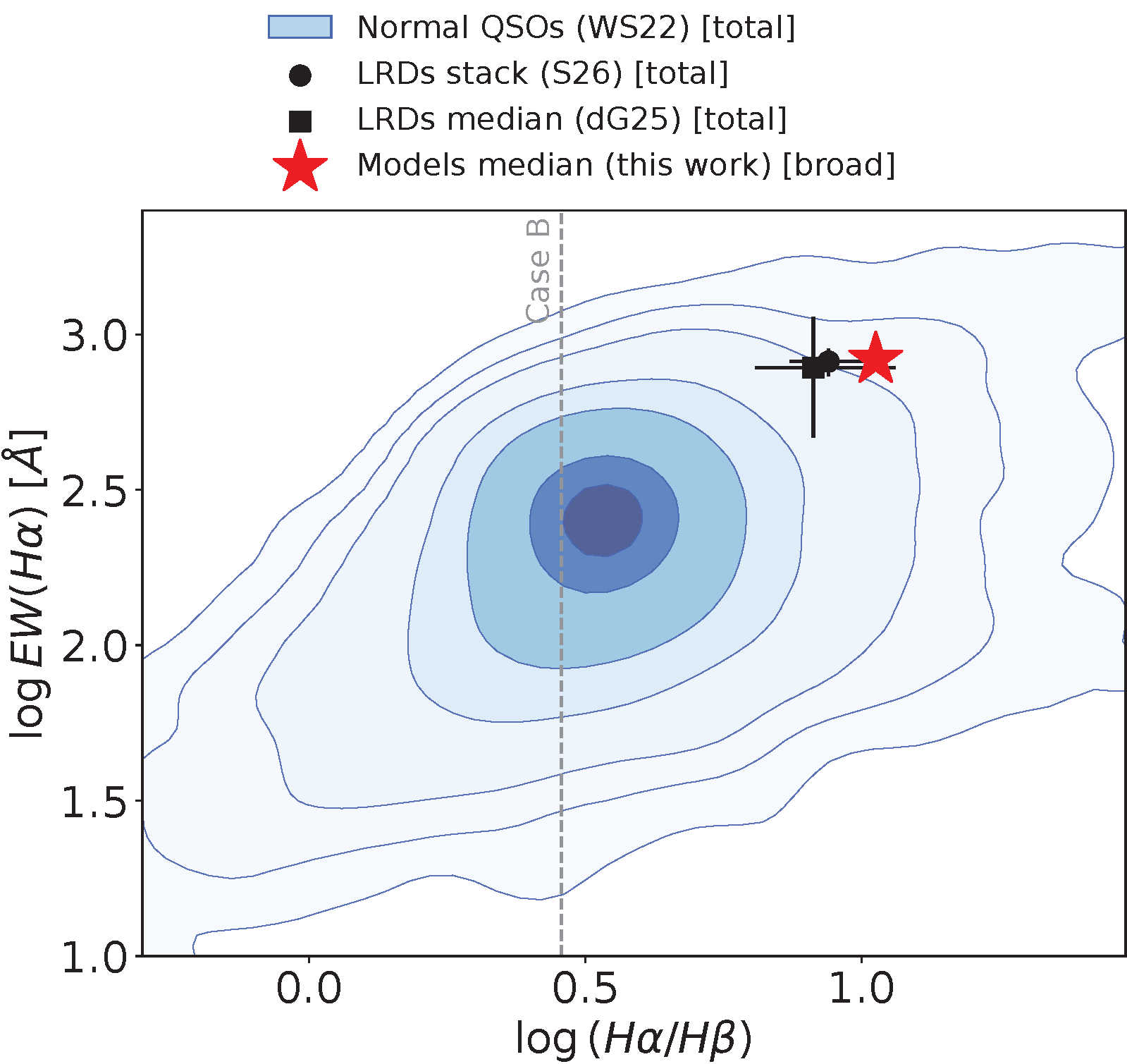}
\caption{Rest-frame H$\alpha$ EW versus Balmer decrement. The red star marks the value from our super-Eddington model matched to the stacked LRD spectrum (broad H$\alpha$ component). Filled symbols show literature measurements for JWST LRDs: the median of the \citet{degraaf2025_multipleBHstars} LRD sample and the stacked LRD spectrum of \citet{Sun2026}. For the de Graaff et al.\ point, the horizontal and vertical error bars denote the 16th and 84th percentiles of the observed distributions in ${\rm H}\alpha/{\rm H}\beta$ and EW(H$\alpha$), respectively. Gray contours show the locus of $z\simeq 0$--0.6 SDSS broad-line quasars from the DR16Q compilation of \citet{Wu2022}. The dotted line indicates the Case~B recombination value $({\rm H}\alpha/{\rm H}\beta)=2.86$. Our model lies on an extreme but continuous extension of the trend defined by classical AGNs toward larger EW(H$\alpha$) and Balmer decrements.
}
\label{fig:EW_BD}
\end{figure}

A direct observational corollary of the dusty, inclination-dependent component in our model is a large Balmer decrement: extinction suppresses H$\beta$ more strongly than H$\alpha$, so even a moderately elevated intrinsic broad-line ratio $(\mathrm{H}\alpha/\mathrm{H}\beta)_{\rm int}$ is readily driven to the high values inferred for LRDs along dust-obscured sightlines. For our fiducial BLR spectrum we obtain $(\mathrm{H}\alpha/\mathrm{H}\beta)_{\rm int}\simeq 4.6$; applying the same line-of-sight attenuation, $A_V=3.1$, required by the LRD SED fits naturally yields $(\mathrm{H}\alpha/\mathrm{H}\beta)_{\rm obs}\simeq 10.6$. Figure~\ref{fig:EW_BD} places this result in the plane of EW(H$\alpha$) versus ${\rm H}\alpha/{\rm H}\beta$: the model point (red star) lies at the high-EW, high-decrement end of the $z\simeq 0$--0.6 SDSS quasar locus \citep{Wu2022}, and is consistent with current JWST LRD constraints from both the median of the \citet{degraaf2025_multipleBHstars} sample and the stacked LRD spectrum of \citet{Sun2026}. In this sense, our model appears as a continuous, albeit extreme, extension of classical type-1 AGN properties rather than a qualitatively distinct population (c.f. \citealt{Sun2026}). We emphasize that Case~B is not expected to hold exactly in BLR conditions, where collisional processes and line optical depths can elevate the intrinsic decrement (indeed, our fiducial ${\rm H}\alpha/{\rm H}\beta\simeq 4.6$ already reflects such departures), but recent multi-line JWST analyses find that the broad-line ratios in most LRDs remain consistent with dust-reddened recombination, with only rare outliers requiring additional radiative-transfer effects \citep{Nikopoulos2025,Deugenio2025_irony}.

In our orientation-based picture, LRDs represent the high-decrement tail of the BLAGN population: along obscured, near-equatorial sightlines the same dusty, low-ionization gas that reddens the continuum preferentially attenuates H$\beta$ relative to H$\alpha$. By contrast, the bluer, less-obscured LBDs are viewed along relatively clear sightlines, so their Balmer decrements should remain only mildly above Case B, with intermediate objects naturally populating the transition regime.

\subsection{An extreme LRD: The Cliff}
\label{sec:Cliff}

The Cliff exhibits the strongest Balmer break reported to date in any galaxy or AGN, roughly twice that of the previous record holder, A2744-45924 \citep[][see also \citealt{Ivey2026}]{deGraaff2025}, together with an exceptionally large broad-line Balmer decrement. It therefore provides an ideal test of the orientation-based, dense-gas framework developed above at the extreme end of the LRD population.

\begin{table}
\caption{Parameters entering the AGN+BLR+dust model fit to The Cliff (RUBIES-UDS-154183; Figure~\ref{fig:Cliff}).}
\label{tab:cliff}
\centering
\begin{tabular}{lcl}
\hline\hline
Parameter & Value & Comment \\
\hline
\multicolumn{3}{l}{Fixed: stack values} \\
$M_{\rm BH}$       & $10^{7.5}\,M_\odot$         & fiducial \\
$\dot{m}$          & $32$                        & fiducial \\
$n_{\rm H}$        & $10^{10}\,{\rm cm}^{-3}$   & fiducial \\
$\sigma_{\rm BLR}$ & $0.2$                       & fiducial \\
\hline
\multicolumn{3}{l}{Fixed: modified for The Cliff} \\
$\log U$           & $-2.0$                      & Balmer decrement \\
$N_{\rm H}$        & $10^{24}\,{\rm cm}^{-2}$   & Balmer break \\
$C_{\rm BLR}$      & $0.045$                     & H$\alpha$ EW \\
$i$                & $85^\circ$                  & near-equatorial \\
$P_{\rm esc}$      & $0$                         & optically thick \\
$B$                & $0$                         & SMC-like \\
\hline
\multicolumn{3}{l}{Free} \\
$A_V$              & $2.54$                      & best fit \\
$R_V$              & $2.70$                      & best fit \\
\hline
\multicolumn{3}{l}{Derived} \\
EW(H$\alpha$)      & $1026$\,\AA                & obs.\ $1106\pm55$\,\AA \\
Balmer break       & $5.73$                      & obs.\ $6.9^{+2.8}_{-1.5}$ \\
H$\alpha$/H$\beta$ & $17.1$                      & intrinsic $7.0$ \\
\hline
\multicolumn{3}{l}{Flux normalization} \\
Anchor window & $4000$--$4100$\,\AA & PRISM\tablefootmark{a}\\
\hline
\end{tabular}
\tablefoot{\tablefoottext{a}{The model is normalized to the median continuum flux in the rest-frame 4000--4100\,\AA\ interval, following the convention of
\citet{deGraaff2025}. Observed EW and Balmer-break values are from
\citet{Rusakov2026} and \citet{deGraaff2025}, respectively. Unlike the stack fit (Table~\ref{tab:dustfit}), $\sigma_{\rm dust}$ and
$C_{\rm dust}$ are not used here because the inclination $i$ and escape probability $P_{\rm esc}$ are fixed directly.}}
\end{table}

We adopt the same physical framework as for the composite LRD SED (Table~\ref{tab:dustfit}), retaining the stack-fit values of $M_{\rm BH}$, $\dot{m}$, $n_{\rm H}=10^{10}\,{\rm cm}^{-3}$, and $\sigma_{\rm BLR}=0.2$. Four parameters are modified to represent a single, nearly edge-on system rather than a population average. The BLR column density is increased to $N_{\rm H}=10^{24}\,{\rm cm}^{-2}$, following the dense-gas model proposed to account for the unusually deep Balmer break of this object \citep{deGraaff2025}; the BLR covering factor is reduced to $C_{\rm BLR}=0.045$ to provide a better match to the observed broad H$\alpha$ equivalent width \citep{Rusakov2026}; and the ionization parameter is lowered to $\log U=-2.0$, consistent with the large Balmer decrement. We do not fit these BLR structural parameters simultaneously, as their effects on the available line and continuum diagnostics are partially degenerate. Instead, we fix $N_{\rm H}$ and $C_{\rm BLR}$ as described above, retain the stack-fit value of $n_{\rm H}$, and restrict the formal $\chi^2$ minimization to the two well-constrained foreground dust parameters, $A_V$ and $R_V$.

Finally, we fix the observer inclination at $i=85^\circ$ and set $P_{\rm esc}=0$. For $\sigma_{\rm BLR}\lesssim0.2$, the cloud distribution (Eq.~\ref{eq:Nlos}) already implies intercepted sightlines with median inclinations exceeding $80^\circ$, while $P_{\rm esc}=0$ enforces the fully optically thick, direct-disk-light-free limit appropriate for the reddest LRDs \citep{Inayoshi2025,degraaf2025_multipleBHstars} and removes the otherwise strong degeneracy between $P_{\rm esc}$ and $A_V$. We likewise fix the UV bump strength at $B=0$, corresponding to a bump-free, SMC-like attenuation curve. This is both the solution preferred by an unconstrained fit to The Cliff photometry and the attenuation curve adopted in the independent SED fit of The Cliff by \citet{Barro2025}. An SMC-like curve is commonly adopted for low-metallicity galaxies \citep[e.g.][]{Reddy2018}; The Cliff itself has $Z\simeq0.017\,Z_\odot$ \citep{Ivey2026}.

\begin{figure}[!hbt]
\centering
\includegraphics[width=\hsize,trim=0 0 0 0,clip]{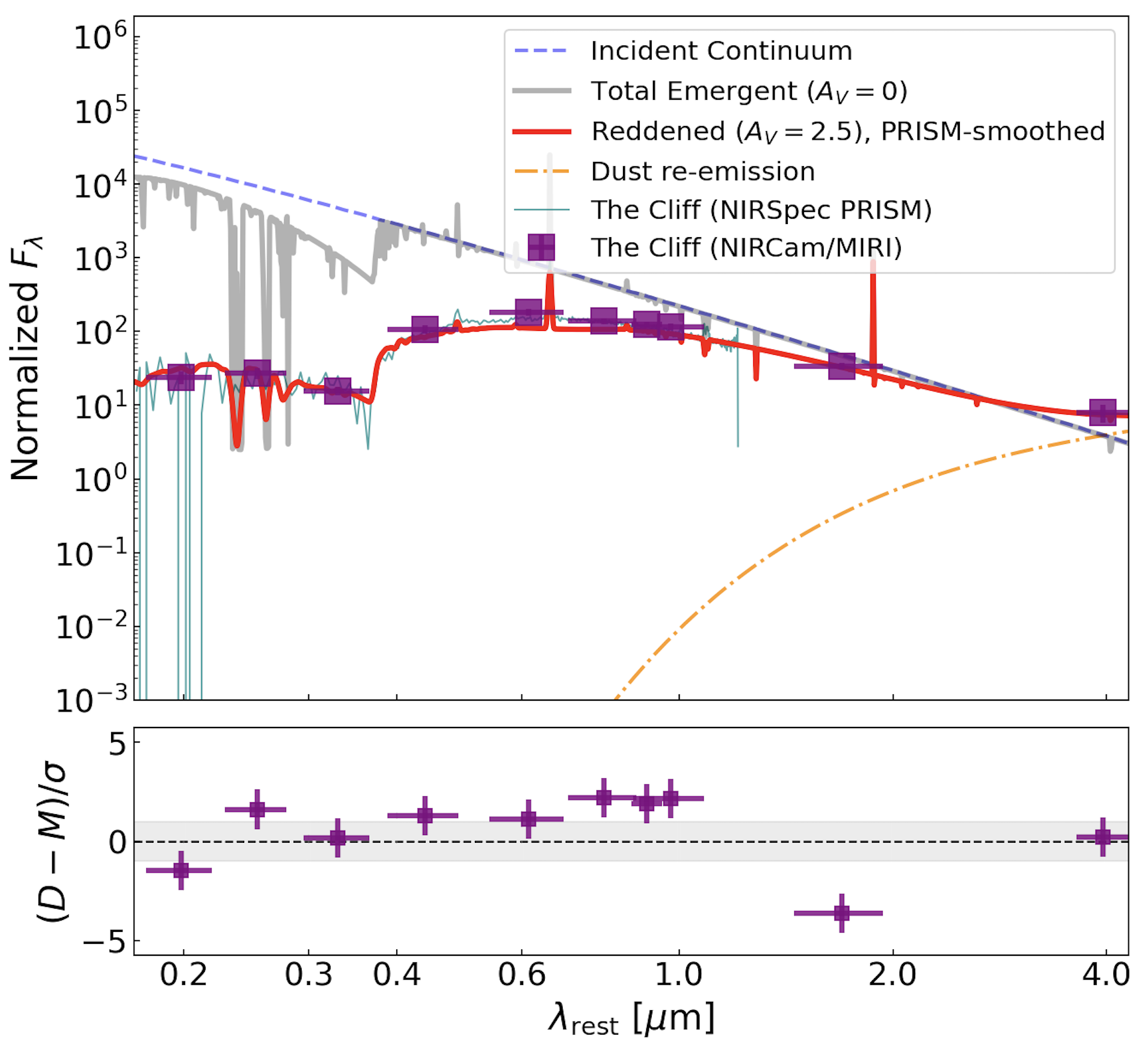}
\caption{Best-fit AGN+BLR+dust model for {The Cliff}
(RUBIES-UDS-154183). {Top:} rest-frame spectral energy distribution, normalized to the PRISM anchor window (see text). The dashed blue line shows the intrinsic incident continuum; the gray curve is the total emergent spectrum (transmitted plus nebular components) prior to dust
attenuation; the solid red curve is this same spectrum after foreground reddening, convolved to the wavelength-dependent NIRSpec PRISM resolution for direct comparison with the data. The dot-dashed orange curve shows an additional dust re-emission (torus) component, described in
\S~\ref{sec:dust}, which is negligible at all wavelengths shown except the rest-frame $\sim\!4\,\mu$m point (observed MIRI/F1800W). The teal line and purple squares show the observed NIRSpec/PRISM spectrum and
NIRCam/MIRI photometry of The Cliff, respectively. {Bottom:}
photometric residuals, $(D-M)/\sigma$, in units of the data uncertainties; the shaded band marks $\pm1\sigma$. The rest-frame $4\,\mu$m point is
not included in the formal $\chi^2$ minimization but is shown for comparison, and is well reproduced once the dust re-emission component is included.}
\label{fig:Cliff}
\end{figure}

With $M_{\rm BH}$, $\dot{m}$, $\log U$, $n_{\rm H}$, $N_{\rm H}$, $C_{\rm BLR}$, $\sigma_{\rm BLR}$, $i$, $P_{\rm esc}$, and $B$ all fixed,
the fit to The Cliff's photometry has only $A_V$ and $R_V$ as free parameters, listed in Table~\ref{tab:cliff}. The best fit, shown in
Figure~\ref{fig:Cliff}, gives $A_V=2.54$ and $R_V=2.70$ ($\chi^2=33.4$ for 7 degrees of freedom, excluding the rest-frame $4\,\mu$m MIRI
point from the fit itself; see \S~\ref{sec:dust}). The model reproduces the observed Balmer break strength within its observational uncertainty, $5.73$ versus $6.9^{+2.8}_{-1.5}$ \citep{deGraaff2025}, and the exceptionally large
broad-line Balmer decrement, predicting ${\rm H}\alpha/{\rm H}\beta=17.1$ after reddening from an intrinsic, density-driven value of $7.0$, which is the average value observed in LBDs \citep{Geris2026} -- this is a
decrement far in excess of the Case~B value that arises naturally from the same dense, optically thick gas responsible for the Balmer break, without invoking implausibly large foreground dust columns. The best-fit model also reproduces the broad H$\alpha$ equivalent width, ${\rm EW}=1026$\,\AA, consistent with the measured value of $1106\pm55$\,\AA\ reported by \citet{Rusakov2026}.

Although the $4\,\mu$m rest-frame photometric point is not included in the fit, its observed flux is naturally explained by the hot dust emission predicted by the model, with the same dust covering factor of 20\% inferred for the average LRD population.

Overall, it is remarkable that the observed fluxes are matched to within
$\sim\!15$--$20\%$ (Fig.~\ref{fig:Cliff}) despite the relative simplicity of the model. The minor residual variations are likely due to imperfections in the adopted SMC-like attenuation curve or in the intrinsic anisotropic accretion-disk continuum, rather than in the BLR reprocessing model.

The only significant discrepancy is the rest-frame $\sim\!1.7\,\mu$m point (observed MIRI/F770W), which is overpredicted by $\sim\!40\%$ because the Pa$\alpha$ line ($\lambda1.875\,\mu$m) falls within the filter bandpass. The population-averaged BLR line ratios adopted here do not fully reproduce the observed Pa$\alpha$/H$\alpha$ suppression.
The Pa$\alpha$/H$\alpha$ ratio is expected to decrease with increasing $n_{\rm H}$ through the same collisional and radiative-transfer effects responsible for the anomalous Balmer decrement. We do not, however, attempt to re-tune $n_{\rm H}$ to match a single photometric point. This residual therefore most likely reflects the limitations of the present single-density, single-column BLR model rather than of the underlying dense-gas scenario.

\subsection{No dust budget crisis in LRDs}
\label{sec:dust}

A commonly noted tension for dust-reddened interpretations of LRDs is an ``IR-energy (dust budget) crisis'': extinctions of $A_V \gtrsim$ a few magnitudes, when coupled with an effectively enclosing dusty screen (solid angle $\Omega \sim 4\pi$), would intercept a large fraction of the bolometric output and predict strong mid-IR to (sub)mm re-radiation, in tension with the faint stacked or non-detected infrared emission in several samples, including the most luminous systems \citep[e.g.,][]{Chen2025,Setton2025b}.

At the same time, stacking analyses have recently revealed an AGN-heated hot-dust component in the median LRD SED, rising into the rest-frame near-IR, and argued that the dust geometry must leave at least part of the optical/UV continuum and BLR emission unobscured \citep{Delvecchio2025}. Recent Rosetta Stone case studies likewise find clear evidence for AGN-heated hot dust in both the LRD archetype GN--28074 and the LBD archetype GS--3073 \citep{Brazzini2026}. In addition, SED modeling of local LRD analogs indicates that hot (torus) dust can dominate the rest-frame near-IR: \citet{Ji2025z01}, for example, find that the best-fit emission around $\sim 5\,\mu$m is dominated by hot dust from the torus, while \citet{Lin2026} report significant hot-dust emission in local LRDs. In our inclination-based framework this is consistent with a modest global dust covering factor: while dust reradiation is approximately isotropic, only a small fraction $C_{\rm dust}$ of the bolometric luminosity is intercepted and reprocessed, allowing a hot-dust rise in the near--IR without the large IR power expected for fully enshrouding (near-unity covering) models.

We model the infrared reprocessing by dust using an energy-balance calculation. Given an intrinsic (unreddened) incident spectrum $L_{\nu,\rm inc}$, dust at radius $r$ absorbs a fraction of the radiation field and re-emits thermally. {A similar framework, based on an extended dust distribution and thermal re-emission, was recently presented  by \citet{Li2025}. Here, however, rather than solving explicitly for $T_{\rm d}(r)$ from a frequency-dependent radiative-equilibrium equation, we adopt a parametric temperature profile anchored at the sublimation radius and normalize the dust emission by enforcing bolometric energy conservation.} Specifically, we take

\begin{equation}
T_{\rm d}(r)=T_{\rm sub}\left(\frac{r}{r_{\rm in}}\right)^{-1/2},
\qquad r_{\rm in}\le r\le r_{\rm out},
\label{eq:Tdust_profile}
\end{equation}

as expected for radiative equilibrium with approximately gray dust heated by a central source. Here, $r_{\rm in}$ is set by dust sublimation, $T_{\rm d}(r_{\rm in})=T_{\rm sub}$. The absorbed luminosity is computed directly from the difference between the unreddened and reddened continua implied by the adopted extinction curve,

\begin{equation}
L_{\rm abs}=\int_{0}^{\infty} \left[L_{\nu,\rm inc}-L_{\nu,\rm inc}\,10^{-0.4A_{\lambda}}\right]\,{\rm d}\nu,
\label{eq:Labs}
\end{equation}

so that the wavelength dependence of dust opacity enters through $A_{\lambda}$ without requiring explicit specification of the extinction or absorption cross sections. Here $L_{\rm abs}$ should be interpreted as the bolometric power removed from the incident spectrum along a representative dust-obscured line of sight; the total power reprocessed by dust is then reduced by the global covering factor $C_{\rm dust}$.

We assume a power-law radial density profile for the dusty gas,

\begin{equation}
n(r)=n_{0}\left(\frac{r}{r_{\rm in}}\right)^{-q},
\qquad r_{\rm in}\le r\le r_{\rm out},
\label{eq:ndust}
\end{equation}

and construct the (unnormalized) re-emitted spectrum as a superposition of modified blackbodies,

\begin{equation}
L_{\nu,\rm dust} \propto
\kappa_\nu
\int_{r_{\rm in}}^{r_{\rm out}}
B_{\nu} \left[T_{\rm d}(r)\right]\,
n(r)\,4\pi r^{2}\,{\rm d}r,
\label{eq:Ldust_template}
\end{equation}

where we adopt an effective opacity law $\kappa_\nu \propto \nu^{p}$ with $p \simeq 0$. This graybody approximation is motivated by the expected optical thickness of the dusty medium rather than by the intrinsic optical properties of the grains. Given the extreme compactness and high column densities inferred for these sources, the dust is likely distributed in dense, optically thick clumps \citep[e.g.,][]{Nenkova2008}. In this regime, the emission is dominated by the optically thick surfaces of the clouds, while radiative transfer effects and the superposition of clumps spanning a range of temperatures dilute intrinsic compositional spectral signatures. The resulting infrared continuum is therefore well approximated by a graybody with an effective emissivity index $p\simeq0$, even if the individual grains have a steeper intrinsic emissivity ($p\approx2$ in the diffuse ISM).

Finally, we set the normalization by requiring that the integrated dust emission equals a fraction $C_{\rm dust}$ of the absorbed luminosity,

\begin{equation}
\int_{0}^{\infty} L_{\nu,\rm dust}\,{\rm d}\nu \;=\; C_{\rm dust}\,L_{\rm abs}.
\label{eq:Ldust_norm}
\end{equation}

\begin{figure}[!ht]
\centering
\includegraphics[width=\hsize,trim=0 0 0 0,clip]{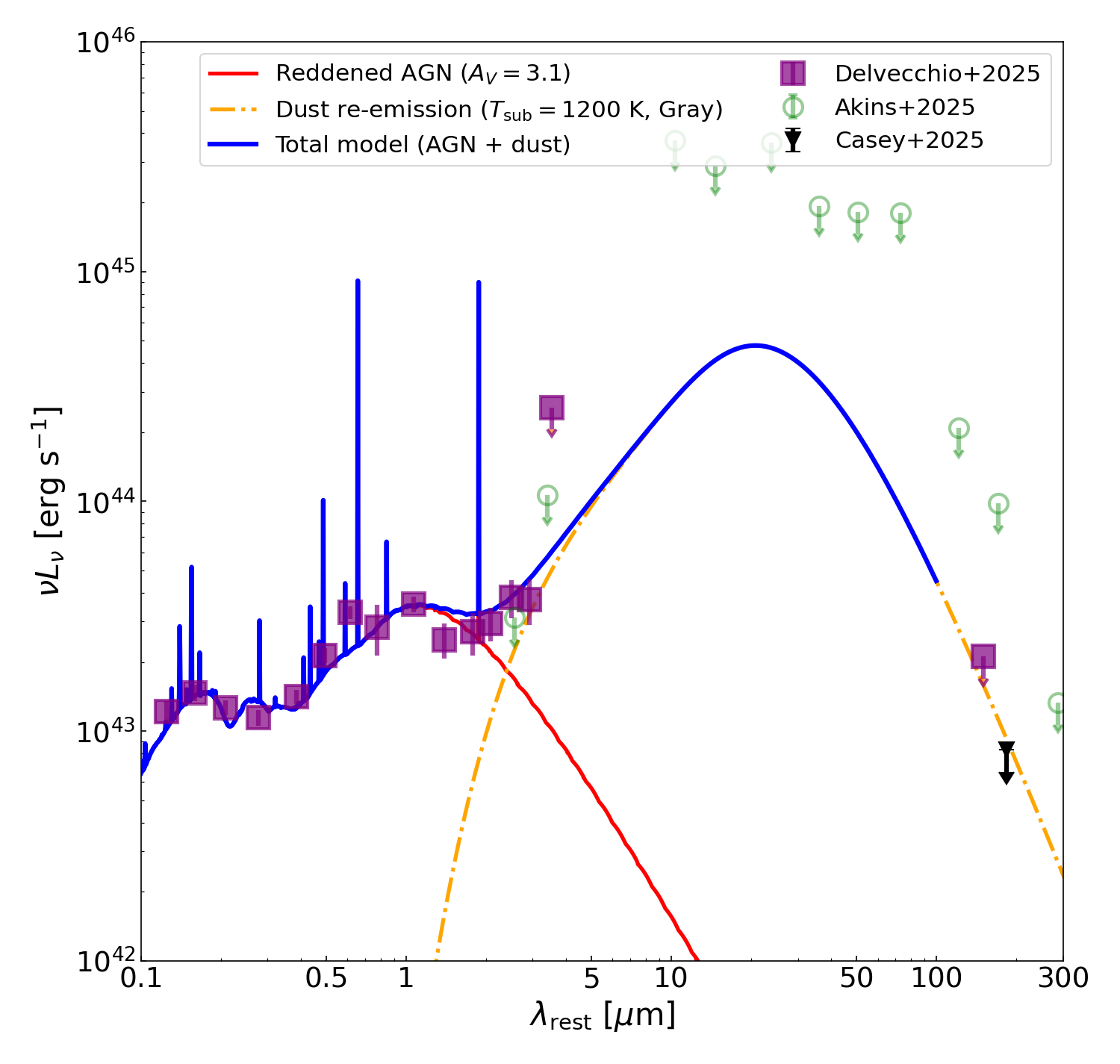}
\caption{Rest-frame $\nu L_{\nu}$ SED illustrating the dust-reprocessed emission in our LRD model. The red curve shows the reddened super-Eddington AGN continuum for a per-cloud extinction $A_V=3.1$, normalized to the Delvecchio et al.\ composite at $\lambda_{\rm rest}\simeq 7800$\,\AA. The dot-dashed orange curve is the dust re-emission computed from the absorbed luminosity using a multi-temperature blackbody with $T_{\rm sub}=1200\,$K, a radial density profile $n\propto r^{-0.3}$, and $r_{\rm out}/r_{\rm in}=100$, scaled by a dust covering factor $C_{\rm dust}=0.2$. The blue curve is the total model (reddened AGN plus dust). Purple squares show the stacked LRD photometry from \citet{Delvecchio2025}, while green open circles and the black arrow indicate the infrared and sub-mm upper limits from the \citet{Akins2025} stack and from \citet{Casey2025}, respectively. The model reproduces the observed near-IR rise of LRDs while remaining consistent with current constraints at $\lambda_{\rm rest}\gtrsim 10\,\mu$m, thereby avoiding a dust-mass or IR-energy budget crisis.
}
\label{fig:dustFIR}
\end{figure}

In the fiducial configuration used in Figure \ref{fig:dustFIR} we adopt $q=0.3$, a sublimation temperature $T_{\rm sub}=1200\,{\rm K}$, and a radial extent $r_{\rm out}/r_{\rm in}=100$. These choices produce a broad mid-IR  bump peaking at $\lambda_{\rm rest}\sim 20\,\mu{\rm m}$ that matches the Delvecchio et al.\ composite while keeping the far-infrared tail consistent with current limits, and they lie within the range of slopes and radial extents commonly inferred for AGN dusty tori in clumpy and smooth models \citep[e.g.,][]{Fritz2006,Nenkova2008,Stalevski2012}.

Although our dust model is not explicitly parameterized in terms of a total dust mass, the implied $M_{\rm dust}$ is very small. For the extinctions relevant to LRDs ($A_V\simeq3$ along obscured sightlines), and assuming $D/G\propto Z$ with $Z\sim0.1\,Z_\odot$, the reduced dust-to-gas ratio ($D/G\simeq10^{-3}$) implies a gas column of $N_{\rm H}\sim(5\times10^{22}$--$10^{23})\,{\rm cm^{-2}}$. For our fiducial compact structure with $C_{\rm dust}=0.20$, $r_{\rm in}\sim0.15\,{\rm pc}$, $r_{\rm out}/r_{\rm in}=100$, and $n\propto r^{-q}$ with $q=0.3$, this corresponds to a total gas mass of several $\times10^{4}$--$10^{5}\,M_\odot$ and hence $M_{\rm dust}\sim60$--$120\,M_\odot$. This is orders of magnitude below the ALMA stacking upper limits for typical LRDs ($M_{\rm dust}\lesssim10^{6}\,M_\odot$ at $z\sim6$; \citealt{Casey2025,Chen2025,Setton2025b}) and is therefore fully consistent with current constraints on the dust budget at high redshift \citep{Chen2025}.

{Our treatment nevertheless remains idealized in two respects. First, we assume a fixed covering factor $C_{\rm dust}$ rather than allowing it to vary with AGN luminosity, as suggested by receding-torus evidence and models \citep[e.g.][]{MadauMaiolinoLF,Maiolino2007,Arshakian2005,Matt2019}. Second, we attribute the fitted extinction entirely to the compact dust distribution responsible for the infrared reprocessing, neglecting any additional reddening from cooler dust in the host galaxy. Incorporating a luminosity-dependent covering factor together with a separate host-galaxy dust component would provide a more self-consistent description and should be explored in future work.}

\section{Discussion and summary}
\label{sec:summary}

The physical nature of LRDs remains debated.  Purely stellar interpretations invoke compact, evolved stellar populations with strong nebular emission to explain the V--shaped SEDs \citep[e.g.,][]{Baggen2024,Perez2024,Leung2025,Labbe2025,Hainline2025}, but several lines of evidence increasingly favor an AGN--dominated optical continuum in at least a substantial fraction of LRDs: the prevalence of broad Balmer lines and extreme Balmer EWs \citep{Hviding2025,Yan2025}, the frequent dominance of an unresolved rest--optical component \citep{Hviding2025,Killi2024}, and spectroscopic signatures around the Balmer break that are more naturally explained by absorption and radiative transfer in high-density gas than by an old stellar continuum \citep{Juod2024JADES,Ji2025,Inayoshi2025,Deugenio2025,Yan2025}.  

Motivated by this empirical picture and by the emerging LBD/LRD taxonomy, we have explored a unified AGN framework in which LRDs and LBDs share a common, super-Eddington central engine and differ
primarily by orientation and line--of--sight obscuration. At $z\gtrsim 6$ the nuclei of galaxies are typically extremely gas rich and rapidly fed by cold inflows and violent disk instabilities, while the black holes are still comparatively small; at fixed physical supply rate $\dot M$, the dimensionless accretion rate $\dot m$ is therefore naturally large, so early growth is expected to proceed through frequent or sustained super-Eddington episodes in many models \citep[e.g.,][]{Madau2014,Lupi2016,Husko2025}. 

\paragraph{H$\alpha$ EW.}

A first key result concerns the extreme broad H$\alpha$ EWs of LRDs.  Using super-Eddington thick--disk SEDs, we have constructed BLR photoionization models that include anisotropic illumination and an inclination--dependent continuum decomposition.  When the inclination distribution is folded in, matching the probability--weighted median H$\alpha$ EW of the JWST BLAGN sample \citep[$\simeq 570\,$\AA;][]{Maiolino2025} requires global BLR covering factors of only $C_{\rm BLR}\simeq 0.1$ (see also \citealt{Madau2026}).  These values are fully consistent with standard estimates for low--$z$ Type~1 AGNs \citep[e.g.,][]{Peterson2006,Pandey2023} and stand in contrast to models that invoke nearly $4\pi$ coverage by dense gas in order to reproduce the same EWs with canonical quasar SEDs \citep{Yan2025}.  In our framework, the large EWs arise from two coupled effects: a harder, ionizing-rich intrinsic SED that enhances the line production efficiency, and an anisotropic radiation field in which high--inclination sightlines suffer strong continuum dilution as the hot inner funnel becomes self--shadowed. The most extreme LRDs therefore correspond to high--inclination views of modest--covering BLRs, rather than to enshrouded, unity--covering geometries. We predict that less--reddened LBDs should preferentially exhibit lower EWs, whereas obscured, high--inclination LRDs should populate the high--EW tail.

{While this paper was under review, \citet{Geris2026} presented a new analysis of the H$\alpha$ EW distributions of LBD and LRD subsamples. Geris et al.  find that LRDs indeed have a substantially higher median broad H$\alpha$ equivalent width than LBDs (563 vs.\ 304\,\AA, a factor of $\sim1.9$), despite the two populations showing comparable redshift distributions. We note that the median broad H$\alpha$ EW measured for the LRD
population by \citet{Geris2026} is somewhat lower than found for
the LRD stack of \citet{Sun2026}, a difference the authors themselves highlight in their comparison of the two stacks. Within our framework, such a difference would naturally correspond to a lower BLR covering factor, $C_{\rm BLR}$, than assumed in our fiducial fit. We defer a detailed modeling of the full EW and Balmer-decrement distributions of the \citet{Geris2026} LRD and LBD samples to future work.
}

\paragraph{High-ionization lines.}

Second, the model accounts for the lack of strong high--ionization UV lines in many LRDs and related JWST BLAGNs \citep{Madau2026}. Because the XUV band that controls $\mathrm{He}^{+}$ and higher ions is strongly suppressed toward equatorial sightlines, the predicted He\textsc{ii}\,$\lambda4686$/H$\beta$ ratios fall well below the upper limits in systems such as Abell~2744--QSO1 for plausible combinations of $M_{\rm BH}$, $\dot m$, and ionization parameter.  In this view, the ``missing hard photons'' problem of LRDs is an outcome of orientation--dependent spectral shaping in super-Eddington funnels, rather than evidence for an extremely dense, fully enclosing neutral envelope. In the same scenario X-ray emission is intrinsically suppressed by Compton-cooling of the corona.

\paragraph{V-shaped SED.}

Third, an intrinsically blue
super-Eddington SED combined with modest equatorial dust reddening reproduces the characteristic V--shaped UV--optical SEDs of LRDs.  Adopting a modified \citet{Cardelli1989} extinction law with $R_V=4.1$ and a strongly suppressed 2175\,\AA\ bump ($B=0.2$) yields an effective curve similar to the flatter, weak--bump extinction laws inferred for high--$z$ galaxies, while remaining steeper in the far--UV than the very gray AGN curve of \citet{Gaskell2004}.  For a per-cloud extinction $A_V\simeq 3$ along obscured sightlines, our fiducial model reproduces well the stacked LRD SED of \citet{Delvecchio2025} from rest--frame $1250\,$\AA\ to $1.4\,\mu$m.

{We caution, however, that a single dust law is unlikely to hold across the full LRD population, and individual objects can require significantly different attenuation properties. The Cliff, at $z=3.548$, lies at a lower redshift than the composite stack and provides a case in point: its best-fit SED instead favors a steeper, essentially bump-free ($B=0$), SMC-like attenuation curve with $R_V=2.7$, consistent with its very low inferred metallicity and with the independent SMC-law fit obtained by \citet{Barro2025} for the same object. This spread -- from a flatter, weak-bump curve for the population-averaged stack to a steeper, bump-free curve for this individual, extreme LRD -- suggests genuine object-to-object
variation in dust grain properties (e.g.\ via metallicity-dependent
grain growth/destruction) rather than a single universal LRD
attenuation law, and is a topic we intend to explore in more detail
in future work.
}

It has recently been suggested that the rest-UV component of LRD spectra is primarily associated with the host galaxy, motivated in part by the anti-correlation between EW([O\,\textsc{iii}] $\lambda5007$) and the Balmer break, as well as by the correlation between the [O\,\textsc{iii}] and UV luminosities \citep{degraaf2025_multipleBHstars}. We caution against over-interpreting these trends. Both EW([O\,\textsc{iii}]) and the Balmer break depend on the optical continuum level, while luminosities of different tracers naturally correlate through their common luminosity-distance scaling.

Furthermore, the weak [O\,\textsc{iii}] emission observed in LRDs with the strongest Balmer breaks does not necessarily imply a host-dominated UV continuum. In luminous quasars, EW([O\,\textsc{iii}]) is known to decline with luminosity (the narrow-line Baldwin effect) and may also decrease if the narrow-line region becomes matter bounded. Both effects could be enhanced in the extremely compact environments of LRDs. Moreover, if LRDs are preferentially selected along sightlines with $A_V\sim 3$, flux-limited samples will be biased toward intrinsically more luminous AGN, whose unobscured counterparts would overlap with the quasar population where weak [O\,\textsc{iii}] emission is common. A reduced EW([O\,\textsc{iii}]) can therefore arise without requiring the UV continuum to be dominated by stellar light. A detailed analysis of the NLR physics in these systems is beyond the scope of this paper and is deferred to future work.

The finding of spatially offset blue components is more compelling evidence for host-galaxy contributions \citep[e.g.,][]{Baggen26_LRDoffsets}; however, while a substantial fraction of LRDs show offset blue light, only a subset are sufficiently close to contaminate the spectrum extracted at the LRD position, whereas many of the remaining offsets may be more relevant for Lyman--Werner irradiation than for the observed UV continuum. Moreover, in a number of systems the UV and optical components are coincident, including strongly lensed cases \citep[e.g.,][]{Furtak2024}, and in others the spectral diagnostics unambiguously associate the UV component with the AGN \citep[e.g.,][]{Labbe2024_Monster}. It is plausible that in some LRDs the UV may be host-dominated. In that case, our scenario would still apply to the optical continuum and line emission, particularly if extinction is sufficiently high (or the attenuation curve sufficiently steep) to suppress the AGN contribution in the UV, as may be the case for the LRDs with very weak UV emission \citep[e.g.,][]{Naidu2025}.

\paragraph{Balmer absorption.}

In our picture, the Balmer discontinuity in the processed spectrum
(transmitted plus nebular) is diluted in the median stack by mixing with the direct disk continuum. Pronounced Balmer breaks are therefore expected only when the direct continuum is heavily suppressed and the processed component dominates the observed SED;
a V-shaped UV-optical continuum, by contrast, can arise from wavelength-dependent attenuation and component mixing even when the Balmer edge itself is modest. {This is qualitatively consistent with the relatively small fraction of LRDs exhibiting strong Balmer breaks, and is demonstrated quantitatively for The Cliff in \S~\ref{sec:Cliff}, where reproducing its near-record break requires both a fully suppressed direct continuum ($P_{\rm esc}=0$, i.e.\ a near-equatorial sightline) and a BLR column a full dex above our fiducial stack value.}

The observability of Balmer absorption is regulated by two separable ``gates''. The first is a dust/obscuration gate: in classical low-$z$ unification, equatorial sightlines are typically extinguished enough that the BLR is hidden entirely (Type~2), well beyond the moderate
$A_V\simeq3$ relevant for LRD-like, partially obscured Type~1
sightlines. The second is a gas-physics gate: Balmer absorption
requires a substantial column of dense, partially neutral gas with
an enhanced $n=2$ population in front of the continuum source, with
sufficient covering to imprint troughs.

This gas is naturally available along near-equatorial rays, primarily
in BLR-associated material (and plausibly the torus's molecular/atomic
layers). At low-$z$, these directions are usually rendered inaccessible by the large dust columns producing Type~2 spectra. At high redshift, a reduced dust-to-gas ratio in chemically young nuclei can instead leave some near-equatorial sightlines only moderately extincted, yielding LRD-like spectra in which broad Balmer emission, and in favorable cases Balmer absorption, remain observable --
consistent with the empirical coupling between redder UV slopes and
stronger Balmer breaks \citep{degraaf2025_multipleBHstars}. LBDs, by contrast, correspond to lower-inclination, lightly extincted views,
and are correspondingly less likely to intersect the dense, partially neutral columns Balmer absorption requires.

\paragraph{Balmer decrement.}

The same dusty component that reddens the continuum also predicts large Balmer decrements along obscured sightlines.  For our fiducial BLR model the intrinsic broad--line ratio is
$(\mathrm{H}\alpha/\mathrm{H}\beta)_{\rm int}\simeq 4.6$, which is driven to $(\mathrm{H}\alpha/\mathrm{H}\beta)_{\rm obs}\simeq 10.6$ once the line-of-sight attenuation required by the SED fits ($A_V\simeq3.1$) is applied.  These values are comparable to the high decrements measured in many LRDs and their correlation with UV slope \citep{deGraaff2025}. {The Cliff illustrates how extreme this effect can become for an individual, near-equatorial sightline: with the higher BLR column and lower ionization parameter required to match its exceptional Balmer  (\S~\ref{sec:Cliff}), the same mechanism predicts $(\mathrm{H}\alpha/\mathrm{H}\beta)_{\rm obs}\simeq17.1$, closely matching its observed decrement.}  In our orientation-based unification, LRDs correspond to the high-decrement tail of the BLAGN population, where dusty, low--ionization gas along high-inclination sightlines both reddens the continuum and selectively suppresses H$\beta$; LBDs, viewed along relatively clear sightlines, retain Balmer decrements close to the intrinsic BLR value (which can be Case B or possibly deviating from it because of intrinsic radiative/collisional processes), with intermediate objects populating the transition regime.

\paragraph{Dusty torus.}

Finally, we have addressed the ``IR--energy'' or dust--budget crisis that afflicts heavily obscured interpretations of LRDs.  Extinctions of $A_V\sim 3$ combined with a near--unity dust covering factor would reprocess a large fraction of the bolometric output into the infrared,
in tension with the faint or undetected mid--IR and (sub)mm emission seen in stacks and individual systems \citep[e.g.,][]{Chen2025,Setton2025b,Casey2025}.  Our energy--conserving dust model instead assumes a compact equatorially concentrated structure with a global covering factor of 20\%, a standard power--law radial density profile, a sublimation temperature characteristic of AGN-heated dust, and a finite radial extent of two decades in radius, normalized such that only a modest fraction of the dust-reprocessed luminosity emerges in the near-IR. This configuration produces a broad mid--IR bump peaking at $\lambda_{\rm rest}\sim 20\,\mu$m that matches the Delvecchio et al.\ stack and remains consistent with the Akins et al.\ and Casey et al.\ upper limits at $\lambda_{\rm rest}\gtrsim 10\,\mu$m.  For the modest columns required to achieve $A_V\simeq 3$ along obscured sightlines and a reduced dust-to-gas ratio appropriate for metal-poor hosts, the implied dust mass in our torus is $\lta$ a hundred solar masses, orders of magnitude below current ALMA limits on the global dust content of LRD hosts. In this sense, the small covering, low--mass dusty structure invoked here is a natural way to reconcile the large line--of--sight reddenings with stringent constraints on integrated dust mass and IR luminosity.

\bigskip
Taken together, these results support an interpretation in which LRDs are
not a distinct class of engines, but rather the obscured, high-inclination tail of a
population of compact, super-Eddington BLAGNs, whose less-reddened, more face-on analogs are observed as LBDs. The combination of anisotropic, intrinsically blue SEDs, modest BLR and dust covering factors, and orientation-dependent continuum dilution appears capable of explaining, within a single framework, the defining phenomenology of LRDs: their extreme H$\alpha$ EWs, large Balmer decrements and Balmer breaks, weak high-ionization UV lines, V-shaped continua, muted IR
output, and the generally low level of rest--UV/optical variability expected when the observed continuum is dominated by the geometrically thick outer flow and reprocessed emission.

Exponential Balmer wings have been reported in many LRDs, particularly in the highest--S/N JWST spectra where a double-sided exponential component often provides a statistically preferred fit, prompting suggestions that electron scattering contributes to the line broadening. However, recent analyses have shown that similar non-Gaussian, approximately exponential Balmer wings are not unique to LRDs but are also common in LBDs and even in nearby Type~1 AGNs \citep{Brazzini2026,Wada2026,Laor2006}. {We have recently shown that such wings arise naturally from a radially stratified, virialized BLR, in which the superposition of emission from clouds spanning a range of radii and characteristic velocities produces an approximately exponential profile, without requiring Thomson scattering to be the dominant broadening mechanism \citep{Madau2026Wings}; a similar conclusion has been reached by \cite{Scholtz2026wings}. Electron scattering may still provide a secondary contribution in some sources, but exponential wings alone neither require nor uniquely diagnose an LRD-specific, pc-scale nuclear cocoon \citep[e.g.,][]{Torralba2026,Brazzini2026}.
}

We note that a nearly $4\pi$ gas--enshrouded scenario has been invoked to explain the weak Ly$\alpha$ emission in one LRD at $z=4.4$ \citep{Torralba2026_Lya}, on the grounds that the AGN ionizing continuum cannot escape in this system. However, this inference is based on a single object, and Ly$\alpha$ can be readily attenuated or destroyed by resonant scattering and absorption in neutral gas and dust in the CGM. Moreover, several other LRDs show clear evidence for prominent Ly$\alpha$ emission \citep[e.g.,][]{Geris2026,Tripodi2025a,Morishita2025Lya,Tripodi2025b,Ji2025}.

It has recently been reported that the detection of water absorption in two LRDs supports a scenario in which the observed optical/near--IR emission arises from a cool atmosphere surrounding the black hole \citep{Wang2026water}. We argue instead that water absorption is naturally explained by a warm circumnuclear dusty torus, which is expected to lie along the line of sight for LRDs and can provide physical conditions conducive to molecular survival and formation. Indeed, water absorption is relatively common in AGNs, including in the local Universe, particularly among dust--obscured systems \citep[e.g.,][]{Garcia2026_water,Gonzalez2010_water}. An additional contribution could arise from the host-galaxy stellar population in those two LRDs, which may become more prominent at longer wavelengths. Assessing this possibility would require a dedicated analysis beyond the scope of this paper.

Future tests of our LBD/LRD unification scheme should move beyond single-object comparisons and instead map the joint demographics of high-redshift BLAGNs to test a quantitatively predicted, orientation-based connection between LBDs and LRDs. This requires measuring the LRD/LBD number ratio versus luminosity \citep{MadauMaiolinoLF}, quantifying the incidence of Balmer breaks and the full Balmer-decrement distribution, and establishing how both correlate with UV-optical continuum shape and line EWs. A key requirement is to obtain robust, inclination-aware estimates of $M_{\rm BH}$ and $\lambda_{\rm Edd}$. In particular, current published values of $\lambda_{\rm Edd}$ for LRDs  are typically $\lambda_{\rm Edd}\sim 0.2$--0.5, which may reflect systematics in standard luminosity inference rather than genuinely sub-Eddington accretion. Many studies estimate $L_{\rm bol}$ from broad H$\alpha$ luminosities using locally calibrated scaling relations that implicitly assume minimal internal attenuation and a near-universal quasar SED. If, as in our framework, the same dusty component that reddens the continuum also attenuates the broad Balmer lines, then $L_{\rm H\alpha}$ (and hence $L_{\rm bol}$) can be underestimated by large factors; moreover, if the intrinsic ionizing SED and its anisotropy differ from the local templates underlying these calibrations, the mapping between $L_{\rm H\alpha}$ and $L_{\rm bol}$ becomes orientation dependent. Taken together, dust attenuation and SED- and inclination-driven departures from local $L_{\rm H\alpha}$--$L_{\rm bol}$ conversions may help reconciling existing line-based $M_{\rm BH}$ estimates with substantially higher intrinsic $\lambda_{\rm Edd}$, as required by our model. Standard single-epoch virial methods may also be biased for a flattened BLR, so independent constraints from multi-line and multi-epoch spectroscopy, physically motivated BLR/disk-wind modeling, and bolometric consistency checks will be essential to confirm accretion rates $\dot{m}\gg 1$.

\label{lastpage}
\bibliographystyle{aa}
\bibliography{aa59606-26}

\end{document}